\documentclass[aps,prb,twocolumn,superscriptaddress]{revtex4-1}

\usepackage{graphicx}  
\usepackage{dcolumn}   
\usepackage{bm}        
\usepackage{amssymb}   
\usepackage{amsmath}   
\usepackage{hyperref}  

\newcommand*{\BCSO}{Ba$_3$CoSb$_2$O$_9$} 					
\newcommand*{\abs}[1]{\lvert{#1}\rvert} 					
\newcommand*{\vect}[1]{\bm{#1}}								
\newcommand*{\matr}[1]{\mathbf{#1}} 						
\newcommand*{\half}[1]{\frac{#1}{2}}  				 		
\newcommand*{\third}[1]{\frac{#1}{3}}  				 		
\newcommand*{\ang}{^\circ}									
\newcolumntype{d}{D{.}{.}{-1}}								

\begin{document}

\title{Avoided quasiparticle decay and enhanced excitation continuum in the\\
spin-$\half{1}$ near-Heisenberg triangular antiferromagnet \BCSO}

\author{David~Macdougal}
\affiliation{Clarendon Laboratory, University of Oxford, Parks Road, Oxford, OX1
3PU, United Kingdom}

\author{Stephanie~Williams}
\affiliation{Clarendon Laboratory, University of Oxford, Parks Road, Oxford, OX1
3PU, United Kingdom}

\author{Dharmalingam~Prabhakaran}
\affiliation{Clarendon Laboratory, University of Oxford, Parks Road, Oxford, OX1
3PU, United Kingdom}

\author{Robert~I.~Bewley}
\affiliation{ISIS Pulsed Neutron and Muon Source, Rutherford Appleton
Laboratory, Harwell Campus, Didcot, OX11 0QX, United Kingdom}

\author{David~J.~Voneshen}
\affiliation{ISIS Pulsed Neutron and Muon Source, Rutherford Appleton
Laboratory, Harwell Campus, Didcot, OX11 0QX, United Kingdom}

\author{Radu~Coldea}
\affiliation{Clarendon Laboratory, University of Oxford, Parks Road, Oxford, OX1
3PU, United Kingdom}

\date{\today}

\begin{abstract}
We explore the magnetic excitations of the spin-$\half{1}$ triangular
antiferromagnet \BCSO{} in its $120\ang$ ordered phase using single-crystal
high-resolution inelastic neutron scattering. Sharp magnons with no decay are
observed throughout reciprocal space, with a strongly renormalized dispersion
and multiple soft modes compared to linear spin wave theory. We propose an
empirical parametrization that can quantitatively capture the complete
dispersions in the three-dimensional Brillouin zone and explicitly show that the
dispersion renormalizations have the direct consequence that one$\rightarrow$two
magnon decays are avoided throughout reciprocal space, whereas such decays would
be allowed for the unrenormalized dispersions. At higher energies, we observe a
very strong continuum of excitations with highly-structured intensity
modulations extending up at least $4\times$ the maximum one-magnon energy. The
one-magnon intensities decrease much faster upon increasing energy than
predicted by linear spin wave theory and the higher-energy continuum contains
much more intensity than can be accounted for by a two-magnon cross-section,
suggesting a significant transfer of spectral weight from the high-energy
magnons into the higher-energy continuum states. We attribute the strong
dispersion renormalizations and substantial transfer of spectral weight to
continuum states to the effect of quantum fluctuations and interactions beyond
the spin wave approximation, and make connections to theoretical approaches that
might capture such effects. Finally, through measurements in a strong applied
magnetic field, we find evidence for magnetic domains with opposite senses for
the spin rotation in the $120\ang$ ordered ground state, as expected in the
absence of Dzyaloshinskii-Moriya interactions, when the sense of spin rotation
is selected via spontaneous symmetry breaking.
\end{abstract}

\maketitle

\section{Introduction}
\label{sec:introduction}

Triangular lattice quantum antiferromagnets have been much studied theoretically
as potential hosts for frustration-enhanced cooperative quantum effects, from
the one-third magnetization plateau phase in applied field protected by a
zero-point quantum gap,\cite{Chubukov1991,Honecker1999,Alicea2009,Farnell2009}
to strongly-renormalized magnon dispersions from non-linear
effects,\cite{Zheng2006,Starykh2006} to conceptual models of quantum spin liquid
phases.\cite{Anderson1973,Kalmeyer1987,Balents2010} While it is well-established
that the nearest-neighbor triangular lattice Heisenberg antiferromagnet (TLHAF)
has non-collinear $120\ang$ magnetic order in the ground
state,\cite{Huse1988,Jolicoeur1989,Singh1992,Bernu1994,White2007} as expected at
the mean-field level, but with a reduced ordered moment, less is known about the
full energy spectrum and in particular about the quantitative description of the
intermediate- to high-energy excitations. Higher-order spin wave theory (SWT)
highlights that the non-collinear order induces strong non-linear effects and
couplings between longitudinal and transverse fluctuations, and as a consequence
magnon dispersions are expected to be strongly downwards renormalized with soft
roton-like minima near the M points (mid-edges of the hexagonal Brillouin zone)
compared to the linear spin wave treatment
(LSWT).\cite{Starykh2006,Chernyshev2009} Such effects are also predicted by
series expansion calculations,\cite{Zheng2006} and indeed experimental evidence
has been reported for roton-like minima and also for SWT-predicted finite magnon
lifetime effects near the top of the dispersion in the spin-$2$ TLHAF
LuMnO$_3$.\cite{Oh2013}

Yet to be experimentally tested quantitatively is a SWT prediction that for the
extreme quantum limit of spin-$\half{1}$, magnons should decay over very large
regions of reciprocal space,\cite{Chernyshev2009,Mourigal2013} with an
alternative scenario proposed by DMRG\cite{Verresen2019} and supported by
dynamical variational Monte Carlo calculations\cite{Ferrari2019} proposing
instead avoided quasiparticle decay due to strong quantum interactions that push
the magnon dispersions below the continuum states. Another important unresolved
aspect is the nature of the high-energy excitations beyond one-magnon energies
and to what extent those could be captured quantitatively by two-magnon
excitations within a spin wave expansion. Alternative approaches propose instead
that the higher-energy continuum excitations are better understood in terms of
pairs of unbound spin-$\half{1}$ spinons,\cite{Mezio2011,Ghioldi2015} with the
magnons at low energies corresponding to two-spinon bound
states.\cite{Ghioldi2018,Zhang2019,Ferrari2019}

\begin{figure*}[tb]
\centering
\includegraphics[width=\linewidth,keepaspectratio]{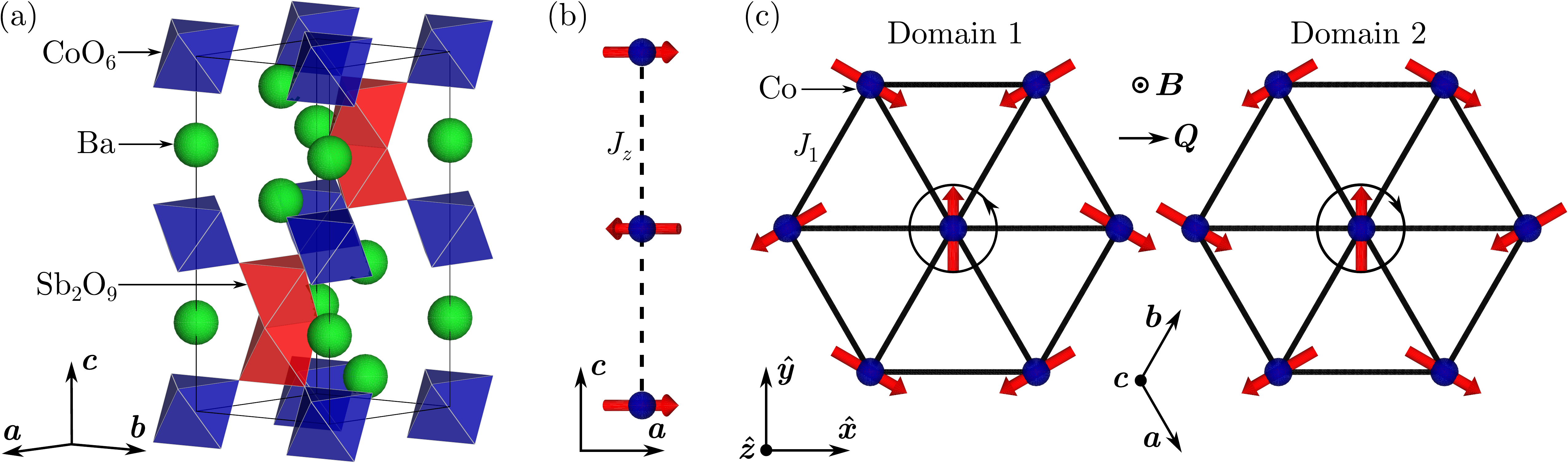}
\caption{(Color online)
(a) Crystal structure of \BCSO{} showing CoO$_6$ octahedra (blue) in the $ab$
plane separated by Ba$^{2+}$ ions (green) and Sb$_2$O$_9$ double octahedra
(red). There are two triangular CoO$_6$ layers in the structural hexagonal unit
cell (thin outline) related by a two-fold screw axis around $c$. (b) Co$^{2+}$ 
spins (red arrows) are ordered antiferromagnetically along $c$ due to the
interlayer exchange $J_z$ (dashed line). (c) In the basal layers the ordered
spins form a non-collinear $120\ang$ structure and two possible magnetic domains
are illustrated here. In domain 1 (left panel), the equilibrium spin direction
rotates counterclockwise (sense indicated by the arrow on the circular envelope
around the central site) between sites whose coordinate along the horizontal
axis increases from left to right (along $\vect{a}+\vect{b}$) when viewed from
above ($\vect{c}$ is out-of-page). In domain 2 (right panel), obtained from
domain 1 by inversion on the central site, the rotation is in the opposite
sense. Thick lines show the nearest neighbor $J_1$ exchange path. The in-plane
projection of the propagation vector $\vect{Q}$ and the direction of the
magnetic field $\vect{B}\parallel\vect{c}$ (for Section~\ref{sec:finite_field})
are also indicated. The diagrams were produced using
\textsc{vesta}.\cite{Momma2011}
\label{fig:Structure}}
\end{figure*}

Motivated by these open theoretical questions, we have revisited the magnetic
excitations of \BCSO{}, proposed to be one of the best realizations of a
near-ideal spin-$\half{1}$ TLHAF with full three-fold lattice
symmetry.\cite{Doi2004,Shirata2012,Zhou2012} The magnetic Co$^{2+}$ ions are
arranged in stacked triangular layers as per Fig.~\ref{fig:Structure}(a)
(hexagonal space group $P6_3/mmc$ with lattice parameters $a=b=5.835$~\AA\ and
$c=14.448$~\AA\ at 1.7~K). The combined effect of local octahedral crystal field
and spin orbit coupling stabilize a Kramers doublet ground state with pseudospin
$S=\half{1}$.\cite{Abragam1951} Magnetic order occurs below 3.8~K in a
non-collinear $120\ang$ structure [see Figs.~\ref{fig:Structure}(b)--(c)] with
spins confined to the basal plane by a small easy-plane exchange
anisotropy.\cite{Susuki2013,Koutroulakis2015,Quirion2015,Ma2016} The high
symmetry of the crystal structure forbids Dzyaloshinskii-Moriya (DM)
interactions between any pair of Co sites located in the same $ab$ plane or
relatively displaced along the $c$ axis; thus DM interactions are ruled out on
all the bonds that are most likely to carry significant exchange interactions.
High-field measurements observed clear evidence for a one-third magnetization
plateau for fields applied in the basal plane,\cite{Shirata2012,Susuki2013} as
expected for the up-up-down phase stabilized by quantum
fluctuations,\cite{Chubukov1991} a phase also observed in the
spatially-anisotropic system Cs$_2$CuBr$_4$.\cite{Ono2003} Previous INS
measurements in \BCSO{} revealed a strong downwards renormalization of the
magnon dispersion, a pronounced roton-like minimum at the M points, and an
extended scattering continuum at higher energies.\cite{Zhou2012,Ma2016,Ito2017}
While the dispersion relations in the one-third magnetization plateau phase
could be well described by a SWT$+1/S$ treatment for a spin Hamiltonian
including easy-plane exchange anisotropy and interlayer
couplings,\cite{Kamiya2018} the observed zero-field dispersions could
\textit{not} be quantitatively described even after including magnon
interactions at order $1/S$ in SWT,\cite{Ma2016} suggesting that quantum
renormalization effects in zero field are much stronger than in the one-third
plateau phase and are underestimated by a perturbative SWT approach.

A quantitative parametrization of the dispersion relations and knowledge of the
energy and wave vector dependence of the continuum scattering intensity are key
pieces of information required by any theoretical models of the many-body
quantum dynamics. Motivated by this, here we present extensive studies of the
magnetic excitations in large single crystals of \BCSO{}\cite{Prabhakaran2017}
with high-resolution inelastic neutron scattering (INS) measurements spanning
multiple Brillouin zones, which reveal that the high-energy excitation continuum
displays highly-structured intensity modulations in momentum space with rings,
hexagons and triangles apparent at various energies. Below the energy threshold
of the continuum scattering, we observe sharp, resolution-limited magnons with
no decay throughout the extended reciprocal space probed. We propose empirical
wave vector-dependent renormalizations of the LSWT dispersion for a spin
Hamiltonian with easy-plane exchange anisotropy, which allow us to
quantitatively capture all modulations of the experimentally-observed magnon
dispersion relations in the full three-dimensional Brillouin zone.

Our main results compared to previous studies\cite{Zhou2012,Ma2016,Ito2017} are
i) the observation that magnons are sharp and do not decay throughout reciprocal
space, and ii) a quantitative parametrization of the complete magnon dispersion
relations in the full 3D Brillouin zone. For the observed strongly-renormalized
dispersion, we find that one- and two-magnon phase spaces in energy and wave
vector never overlap, so the magnon decays are in fact kinematically disallowed
throughout the Brillouin zone, consistent with the experimental observation of
sharp magnons throughout the probed reciprocal space. We note that while the
absence of magnon decays cannot be understood within a SWT approach for the
spin-$\half{1}$ TLHAF, it could in principle be explained if one assumes
substantial easy-plane exchange anisotropy, which gaps out the primary
one-magnon dispersion at the ordering wave vector and thus reduces very rapidly
the overlap phase space, with no overlap expected for $\Delta\lesssim0.92$
($\Delta=1$ is the Heisenberg exchange limit). However, as pointed out by
previous studies,\cite{Ma2016} the predicted magnon dispersions in this case of
substantial easy-plane anisotropy are not compatible with the
experimentally-observed dispersions. This suggests that quantum interaction
effects between one-magnon and higher-energy continuum states in the actual
material are significantly stronger than can be captured perturbatively by SWT
at the $1/S$ level. This could be consistent with recent density matrix
renormalization group (DMRG) calculations, which proposed avoided quasiparticle
decay due to strong interactions in spin-$\half{1}$ models weakly perturbed away
from the TLHAF limit.\cite{Verresen2019} Furthermore, we also observe direct
evidence for a transfer of spectral weight from the one-magnon states to the
higher-energy continuum, which may be understood (at least phenomenologically)
as a further consequence of such strong interactions.

The rest of this paper is organized as follows. Section~\ref{sec:experiment}
describes the experimental setup used for the single-crystal INS measurements.
The following section (Sec.~\ref{sec:zero_field}) presents the results for the
magnetic excitations in the $120\ang$ ordered state at low temperatures and zero
applied magnetic field, starting in Sec.~\ref{sec:results} with an outline of
the key features of the dispersion relations and the intensity modulations in
the high-energy continuum scattering. Section~\ref{sec:LSWT} reviews LSWT
predictions of the magnon dispersions for a spin Hamiltonian with
nearest-neighbor couplings and easy-plane exchange anisotropy.
Section~\ref{sec:renorm} proposes empirical renormalizations of the analytic
LSWT dispersion that can capture quantitatively the observed magnon dispersions
in the full three-dimensional Brillouin zone and Sec.~\ref{sec:fits} describes
the fits to the INS data. Section~\ref{sec:one2two} verifies that
one$\rightarrow$two magnon decays are kinematically disallowed for the
parametrized one-magnon dispersion relation, thus providing a consistency check
for the observation of sharp magnons with no decay throughout the reciprocal
space probed. Section~\ref{sec:two_magnon} presents a quantitative comparison of
the high-energy continuum scattering lineshapes with a two-magnon cross-section,
highlighting which features can and which cannot be captured by such an
approach. Section~\ref{sec:finite_field} presents INS measurements of the spin
dynamics in the cone phase in a $c$-axis magnetic field; the evolution of the
dispersion relations with increasing field are in good (qualitative) agreement
with a LSWT description when including symmetry-allowed magnetic domains with
opposite senses of spin rotation in the $ab$ plane, as illustrated in
Fig.~\ref{fig:Structure}(c). Finally, conclusions are summarized in
Sec.~\ref{sec:conclusion}. The two appendices contain further technical details
on the analysis. Appendix~\ref{sec:appendix_LSWT} presents LSWT calculations for
the magnon dispersion relations and the one- and two-magnon dynamical structure
factor, and sum rules for the total scattering used in the analysis to relate
one- and two-magnon intensities. Appendix~\ref{sec:appendix_renorm} presents
analytic expressions for the wave vector and energy-dependent renormalizations
used to parametrize the observed magnon dispersions.

\section{Experimental details}
\label{sec:experiment}

The spin dynamics of a sample of two co-aligned single crystals of \BCSO{},
grown via the floating zone technique\cite{Prabhakaran2017} (total mass 4~g),
was measured using the direct-geometry time-of-flight neutron spectrometer LET
at the ISIS neutron source in the UK.\cite{Bewley2011} For the zero-field
measurements,\cite{exp_doi_no_field} the sample was cooled by a
variable-temperature insert with He$^4$ exchange gas. Data were collected both
at a base temperature of $1.7$~K, well below the magnetic ordering transition
near $3.8$~K,\cite{Doi2004,Prabhakaran2017} and at $32$~K, deep in the
paramagnetic phase. The spectrometer was operated in repetition rate
multiplication (RRM) mode to collect the inelastic scattering simultaneously for
monochromatic incident neutrons with energies $E_\mathrm{i}=3.53$ and 7.01~meV,
with energy resolutions on the elastic line of 0.062(1) and 0.159(4)~meV (full
width at half maximum, FWHM), respectively. The first configuration provided
high-resolution measurements of the magnon dispersions, which extend up to
$\sim1.6$~meV, whereas the second configuration probed the higher-energy
scattering continuum extending up to at least 6~meV. The higher $E_\mathrm{i}$
data were normalized to give matching magnetic intensities to the lower
$E_\mathrm{i}$ data in the overlapping region of energy transfers near
$E\simeq2$~meV, where the magnetic signal is a broad continuum in both wave
vector and energy. The sample was mounted with the $c$ axis normal to the
horizontal scattering plane, in order to probe the inelastic scattering in
several Brillouin zones in the $hk0$ plane and (via scattering through the
vertical opening of the magnet windows) access also more than a full Brillouin
zone in the interlayer direction. The inelastic scattering was collected in
Horace scans by rotating the sample around the vertical axis in an angular range
of $140\ang$ in steps of $0.5\ang$. Counting times for each orientation were
15~minutes at the base temperature and 7~minutes in the paramagnetic phase, with
an average proton current of 40~$\mu$A.

The same sample and a similar setup were used to measure the inelastic
scattering in a magnetic field applied along the $c$
axis,\cite{exp_doi_in_field} provided by a vertical 9~T cryomagnet. In this
case, the sample was cooled using a dilution refrigerator and the inelastic
scattering was measured at 3, 6 and 9~T at a base temperature of $0.1$~K. The
spectrometer was operated in RRM mode for incident energies $E_\mathrm{i}=2.24$,
3.81 and 7.83~meV, with resolutions on the elastic line of 0.030(1),  0.064(1)
and 0.179(8)~meV (FWHM), respectively. Data were collected in Horace scans
covering a similar range to zero-field measurements with coarser angular steps
and average counting times of 8~minutes per orientation. The time-of-flight
neutron data were processed using the \textsc{mantid}\cite{Arnold2014} and
\textsc{horace}\cite{Ewings2016} data analysis packages.

In order to maximize the counting statistics, for several of the plots in the
paper the intensities were averaged between pixels from the full
four-dimensional Horace scan with wave vector transfers $\vect{k}$ equivalent
under symmetry operations of the crystal lattice point group ($6/mmm$).
All those operations conserved $\abs{\vect{k}}$, so the intensities of all
averaged pixels had the same (spherical) magnetic form factor.

\begin{figure}
\centering
\includegraphics[width=\linewidth,keepaspectratio]{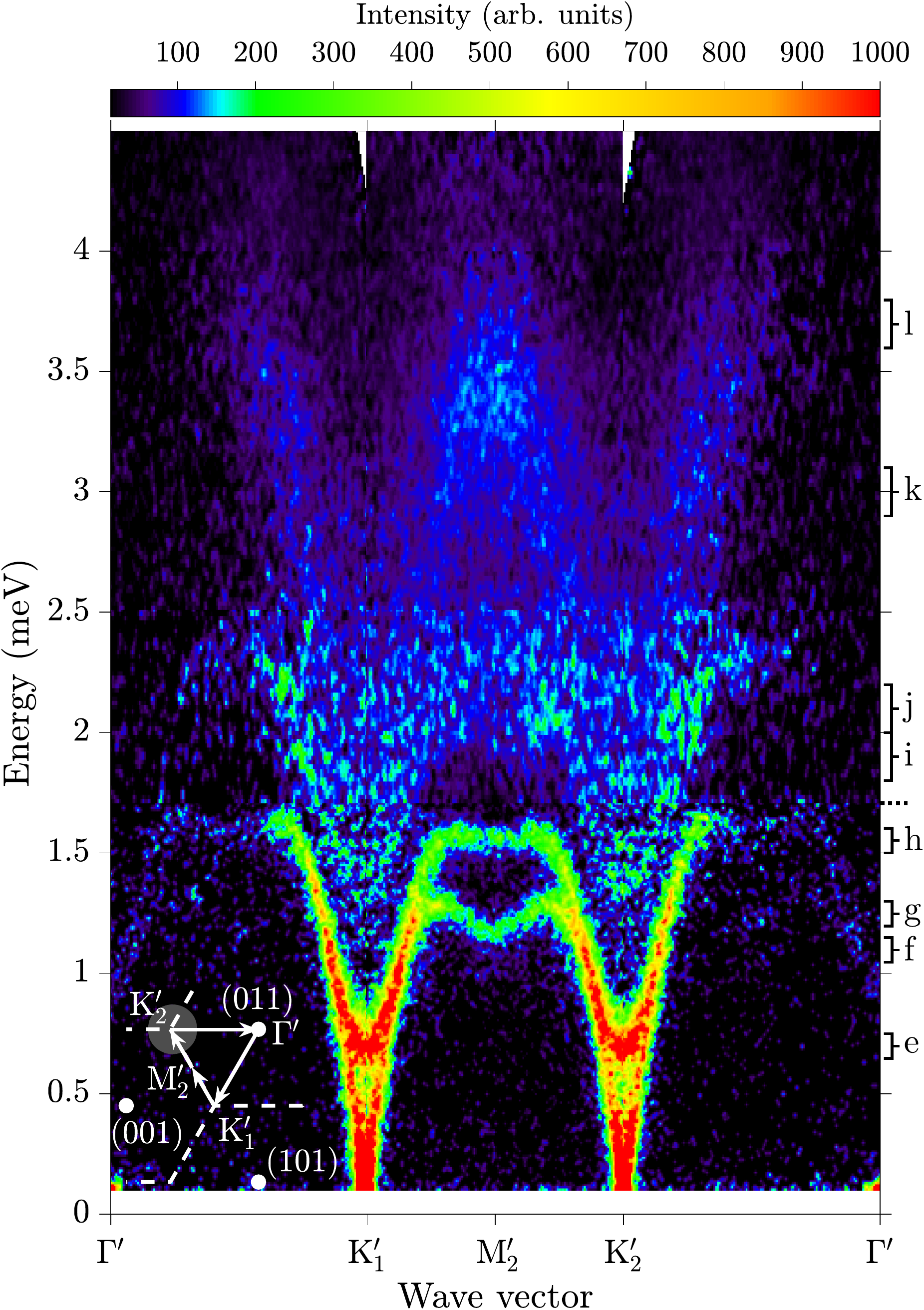}
\caption{(Color online) Observed INS intensity ($1.7$~K) as a function of energy
and wave vector transfer along a high-symmetry path in reciprocal space that
crosses two magnetic Bragg peak positions (K$'_{1,2}$). The color shows the raw
neutron counts in arbitrary units. Two sharp magnon dispersion branches are
clearly observed, accompanied by a strong scattering continuum with a structured
intensity pattern. The brackets on the right-hand side labeled (e)-(l) indicate
the energy integration ranges for the $hk$ slices with the same panel labels
plotted in Fig.~\ref{fig:hkslices}. The wave vector path is $\Gamma' \, (0,1,1)
\rightarrow \mathrm{K}'_1 \, (\third{1},\third{1},1)  \rightarrow \mathrm{M}'_2
\, (0,\half{1},1) \rightarrow \mathrm{K}'_2 \, (-\third{1},\third{2},1)
\rightarrow \Gamma'$, illustrated by arrows in the white bottom-left inset with
the Brillouin zone boundaries represented by dashed lines. The data below
0.1~meV has been omitted as it is dominated by incoherent quasielastic
scattering. The data up to 1.7~meV (horizontal dotted line on the right-hand
side) was collected using $E_\mathrm{i}=3.53$~meV, and at higher energies with
$E_\mathrm{i}=7.01$~meV, scaled as described in the text. The wave vector
integration range in the $hk$ plane is $\pm0.05$~\AA$^{-1}$, and along $l$ is
$\pm0.05$~\AA$^{-1}$ for energies $E\leq2$~meV, $\pm0.1$~\AA$^{-1}$ for
$2<E\leq4$~meV and $\pm0.2$~\AA$^{-1}$ for $E>4$~meV.
\label{fig:KMK_3meV_7meV}}
\end{figure}

\begin{figure*}[t]
\centering
\includegraphics[width=\linewidth,keepaspectratio]{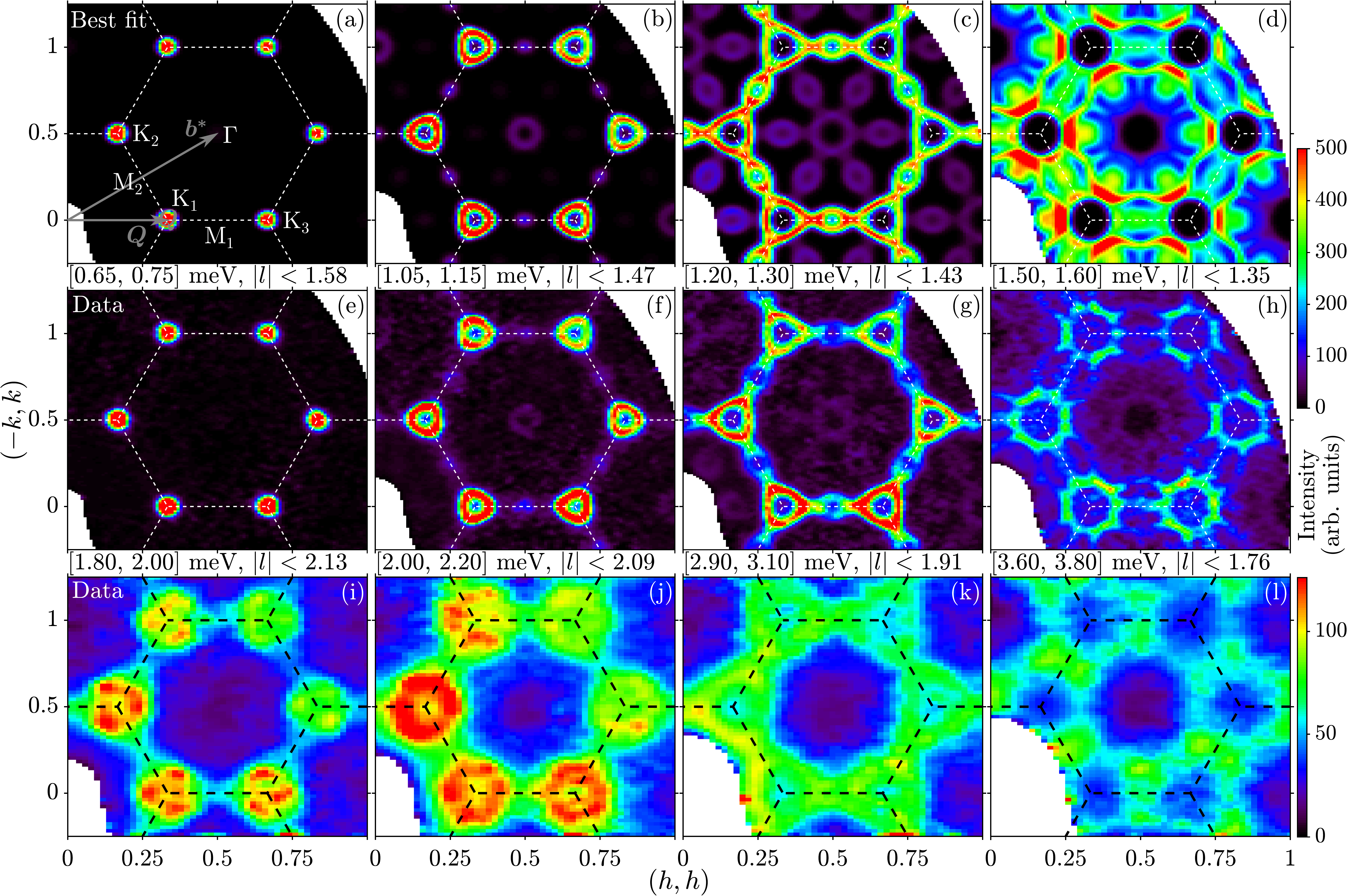}
\caption{(Color online) INS intensity maps as a function of momentum in the $hk$
plane at a series of constant energies, compared with model calculations.
(e)--(h) Intensity maps up to 1.6~meV showing constant-energy contours of the
one-magnon dispersions. (a)--(d) Corresponding calculations for the best-fit
spin wave model with renormalized dispersions described in the text. The model
is the one-magnon cross-section, including the magnetic form factor for
Co$^{2+}$ ions, the neutron polarization factor, the finite-temperature Bose
factor and convolution with the instrumental resolution (for details see
Appendix~\ref{sec:appendix_LSWT}). (i)--(l) Intensity maps through the continuum
scattering at higher energies (above 1.8~meV). The data were collected at 1.7~K
with $E_\mathrm{i}=3.53$~meV for (e)--(h) and 7.01~meV for (i)--(l). In all data
panels, the color shows the raw neutron counts in arbitrary units. In all
panels, dashed lines indicate hexagonal Brillouin zone boundaries. In panel (a),
gray arrows show the projections of $\vect{b}^\ast$ and $\vect{Q}$ wave vectors,
and labels K$_{1\text{--}3}$, M$_{1,2}$ and $\Gamma$ indicate high-symmetry
points referred to in the rest of the paper. The energy integration ranges of
the data panels are given in the panel titles and also indicated by the labeled
brackets on the right of Fig.~\ref{fig:KMK_3meV_7meV}. \label{fig:hkslices}}
\end{figure*}

\section{Spin dynamics in zero field}
\label{sec:zero_field}

\subsection{Key features of the magnon dispersions and continuum scattering}
\label{sec:results}

We begin by presenting the results for the spin dynamics in zero applied field
at a base temperature of $1.7$~K. It is well established
experimentally\cite{Susuki2013,Koutroulakis2015,Quirion2015,Ma2016} that the
magnetic structure in the ground state has spins ordered at $120\ang$ relative
to nearest-neighbor sites in the triangular layers, as illustrated in
Fig.~\ref{fig:Structure}(c), and antiparallel between adjacent layers stacked
along $c$, see Fig.~\ref{fig:Structure}(b). Compared to the structural unit
cell, the magnetic unit cell is tripled in the $ab$ plane, but is the same
length along $c$, with two triangular layers per unit cell. The magnetic
structure can be described in terms of a single propagation vector
$\vect{Q}=(\third{1},\third{1},1)$, where throughout we index wave vectors in
terms of reciprocal lattice units ($h,k,l$) of the hexagonal structural unit
cell. The in-plane components of $\vect{Q}$ capture the $120\ang$ order in a
single layer and the out-of-plane component captures the antiferromagnetic order
between layers spaced by $c/2$. In the absence of DM interactions, the two
senses of spin rotation in the triangular layers [counterclockwise/clockwise
illustrated in Fig.~\ref{fig:Structure}(c) left/right panels] are degenerate, so
one expects a macroscopic sample to contain magnetic domains of both types. In
the absence of bond-dependent spin-exchange anisotropies, believed to be
negligible here, the two magnetic domains have identical excitation spectra in
zero field. (We will show later in Sec.~\ref{sec:finite_field} that the two
domains have different spectra in a finite $c$-axis magnetic field.)

An overview of the observed excitation spectrum as a function of energy and wave
vector is shown in Fig.~\ref{fig:KMK_3meV_7meV} along a representative path in
reciprocal space. Throughout this paper, wave vector labels $\Gamma$, M and K
refer to the conventional high-symmetry points in the two-dimensional (2D)
hexagonal Brillouin zone, where an unprimed (primed) label indicates $l=0$
($l=1$) and numbered subscripts (as in M$_{1,2}$) refer to distinct points in
reciprocal space that are related by a symmetry operation of the lattice point
group when reduced to the first Brillouin zone. Figure~\ref{fig:KMK_3meV_7meV}
shows that the inelastic scattering intensity is strongest near the magnetic
Bragg wave vectors K$_{1,2}'$, and two sharp, well-defined magnon branches are
clearly resolved: one gapless and linearly dispersing at low energies, and the
other one gapped ($E_{\mathrm{gap}}\simeq0.7$~meV) at the magnetic Bragg
position. These modes correspond to the gapless Goldstone mode associated with
rotation of the spins in the $ab$ plane and an out-of-plane mode that is gapped
in the presence of easy-plane (exchange) anisotropy, respectively. In the center
of the figure at the M$'_2$ point, there is a clear local minimum (roton-like
soft mode) in the lower dispersive branch, where the energy is $\sim$8\% lower
compared to that of the nearby local maximum in that branch; a flattening of the
dispersion and a less pronounced soft mode ($\sim$1\% relative dip) is also
visible in the top branch. We will refer to these later as the lower/higher soft
modes, respectively. At the energies of the soft modes, there is almost no
detectable dispersion along the interlayer direction, so we regard these soft
modes as a consequence of the two-dimensional physics in the triangular layers.

Important features of the dispersions are also highlighted in constant-energy
intensity maps. In particular, the triangular-shaped contours with 3-fold
rotational symmetry around the Brillouin zone corners in
Figs.~\ref{fig:hkslices}(f),(g) are characteristic of the spin wave dispersion
shape on the triangular lattice, and the oval-shaped contours near the
mid-points of the zone in Fig.~\ref{fig:hkslices}(g) are due to the soft mode at
M points in the lower magnon mode. Returning to Fig.~\ref{fig:KMK_3meV_7meV},
there is considerable inelastic signal above the sharp magnon dispersions in the
form of a highly-structured continuum, present already inside the magnon
dispersion cones (emerging out of the magnetic Bragg peaks) and extending higher
in energy up to at least 6~meV (data shown up to 4.5~meV in
Fig.~\ref{fig:KMK_3meV_7meV}). The continuum intensity is strongly modulated in
both energy and wave vector. This is clearly illustrated in the intensity maps
at constant energy in Figs.~\ref{fig:hkslices}(i)--(l). At energies just above
the top of the one-magnon dispersions [panel (i)], the continuum intensity is
strongest above the magnon cones centered at K points with a clear 3-fold
symmetric pattern. At slightly higher energies [panel (j)], ring patterns around
K become apparent, and these transform [in panel (k)] into triangular contours
with corners touching at M points. At even higher energies [panel (l)], the
signal near M points has spread out in the direction normal to the Brillouin
zone edges, such that the intensity is strongest along hexagonal contours
centered at $\Gamma$ and connected across M points between adjacent Brillouin
zones. All the above features become overdamped in the paramagnetic phase at
32~K (not shown), confirming their magnetic character.

\subsection{Magnon dispersions within linear spin wave theory}
\label{sec:LSWT}

To parametrize the dispersion relations, following previous
studies\cite{Susuki2013,Ma2016} we consider the minimal spin Hamiltonian
\begin{align}
\mathcal{H}&=J_1\sum^{\mathrm{NN}}_{\langle{ij}\rangle}S^x_iS^x_j+S^y_iS^y_j
+\Delta{}S^z_iS^z_j
\nonumber\\
&+J_z\sum^{\mathrm{interlayer}}_{\langle{mn}\rangle}S^x_mS^x_n+S^y_mS^y_n
+\Delta{}S^z_mS^z_n,
\label{eq:Hamiltonian}
\end{align}
where the nearest-neighbor (NN) intralayer exchange $J_1$, the interlayer
exchange $J_z$ (both antiferromagnetic), and the orientation of the ($x,y,z$)
axes are all illustrated in Figs.~\ref{fig:Structure}(b)--(c). $\Delta<1$
parametrizes the easy-plane exchange anisotropy. This spin Hamiltonian has
continuous rotational symmetry about the $z$ axis in spin space. The crystal
structure however has only discrete rotational symmetries, so we have neglected
in the above Hamiltonian symmetry-allowed bond-dependent exchange anisotropy
terms, such as different exchange couplings for the in-plane spin components
along and perpendicular to a NN bond.

The mean-field ground state of the Hamiltonian in Eq.~(\ref{eq:Hamiltonian}) has
$120\ang$ spin order in the layers and AFM stacking along $c$, as illustrated in
Figs.~\ref{fig:Structure}(b)--(c). The derivation of the dispersion relations
and dynamical structure factor within LSWT is reviewed in
Appendix~\ref{sec:appendix_LSWT}. Three magnon modes are expected for a general
wave vector $\vect{k}$: a primary mode $\omega(\vect{k})$ and two secondary
modes $\omega^{\pm}(\vect{k})\equiv\omega(\vect{k}\pm\vect{Q})$, where
$\vect{Q}$ is the propagation vector of the magnetic structure. For a given wave
vector $\vect{k}$, in general only two out of the three modes carry significant
weight (for the dynamical structure factor calculation see
Appendix~\ref{sec:appendix_LSWT}).

We discuss below the key properties of the primary mode, as the secondary modes
are easily obtained by wave vector translations. The primary mode is gapless at
the origin $\vect{k}=\vect{0}$, corresponding to the Goldstone mode of spin
rotations in the $xy$ plane. For finite easy-plane anisotropy ($\Delta < 1$),
the primary mode has a gap at the magnetic Bragg peak positions
$\vect{k}=\pm\vect{Q}$ of magnitude $E_{\mathrm{gap}}=
3\sqrt{3/2}\:J_1S\sqrt{1-\Delta}$ for $J_z=0$. The interlayer coupling $J_z$
leads to a finite dispersion along $l$ with a zone boundary energy at $(001)$ of
magnitude $6S\sqrt{J_1J_z}$ for $\Delta=1$. Previous
studies\cite{Ma2016,Ito2017} have shown that LSWT for the above spin Hamiltonian
can be used to parametrize well the low-energy dispersions in \BCSO{} up to an
energy of the order of the interlayer zone boundary energy. However the
dispersions at higher energies, in particular close to the top of the
dispersions, could not be accounted for.\cite{Ma2016} Even when including magnon
interaction effects to order $1/S$, the maximum magnon energy was overestimated
by about 45\%, suggesting that quantum renormalization effects on the magnon
dispersions are stronger than can be captured perturbatively at order $1/S$ in
SWT. In the following, to make progress we propose an empirical parametrization
of the dispersion relations.

\begin{figure}[ht]
\centering
\includegraphics[width=\linewidth,keepaspectratio]{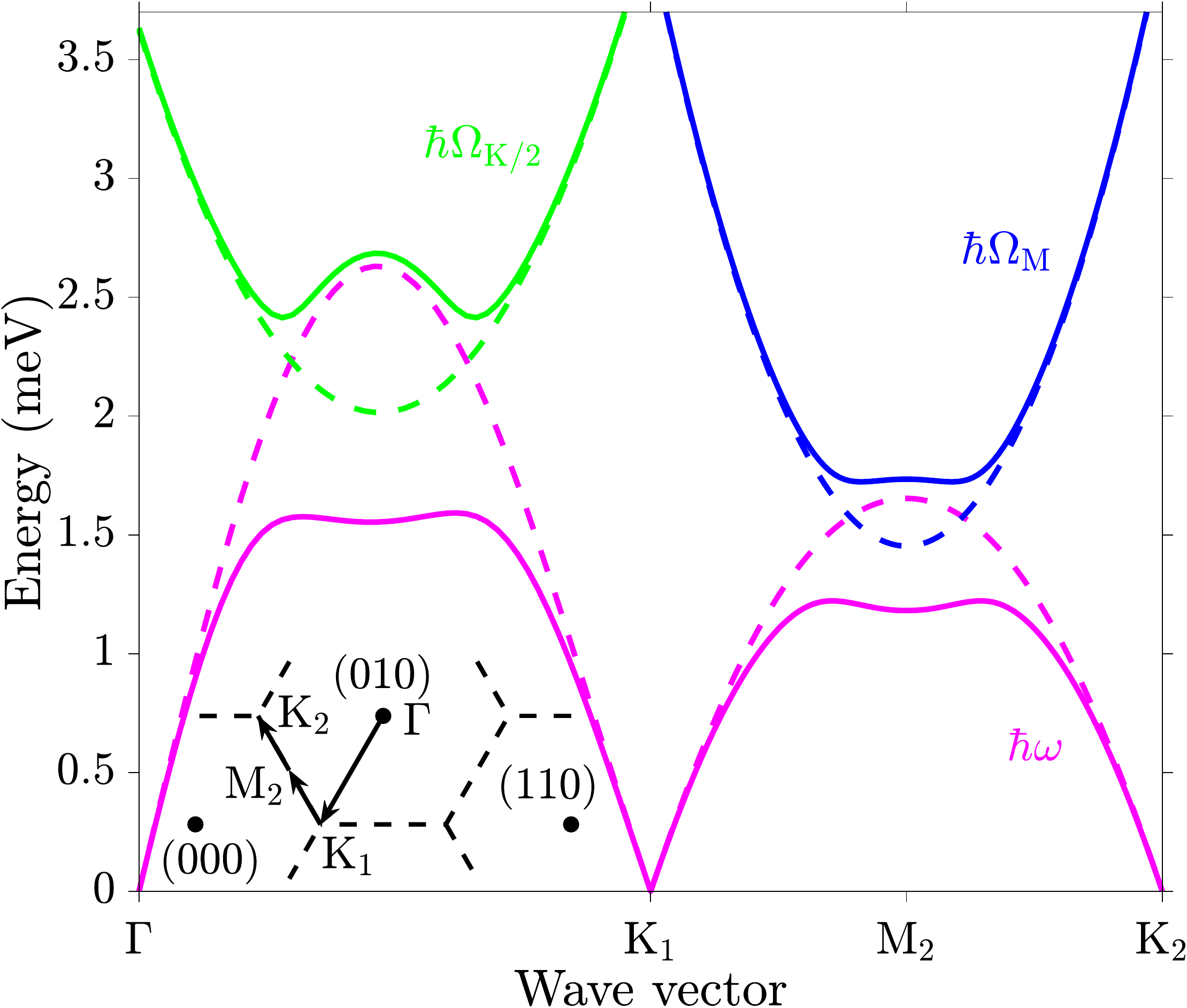}
\caption{(Color online) Illustration of the renormalizations applied to the bare
LSWT dispersion in order to capture the experimental magnon dispersion. For
simplicity, all calculations are for the special case of the isotropic 2D TLHAF
model ($J_z=0, \Delta=1$). Dashed magenta lines show the bare dispersion
$\hbar\omega_\mathrm{LSWT}$ in Eq.~(\ref{eq:Dispersion}) plotted along a
high-symmetry path in the Brillouin zone [schematically shown in the bottom left
inset]. The soft modes at M and near K/2 are introduced by adding a virtual
interaction with fictitious parabolic modes $\hbar\Omega_\mathrm{M}$ (blue
dashed line) and $\hbar\Omega_\mathrm{K/2}$ (green dashed line). The resulting
renormalized dispersion $\hbar\omega$ (solid magenta line) is fitted to the
experimental magnon dispersion.
\label{fig:renorm}}
\end{figure}

\subsection{Proposed empirical parametrization of the observed magnon
dispersions}
\label{sec:renorm}
From general arguments, one expects that the physical magnon dispersion would
satisfy the same periodicity in reciprocal space and the same lattice point
group symmetries as the LSWT dispersion, but that it may be squeezed, stretched
or otherwise deformed compared to the LSWT prediction at various momenta and/or
energies. In this spirit, we introduce below wave vector-dependent
renormalizations that preserve the lattice point group symmetries and allow us
to quantitatively capture all dispersion modulations in the full
three-dimensional Brillouin zone. All operations are performed on the primary
magnon dispersion, as the secondary modes are obtained simply by a wave vector
shift. The complete analytical forms of the renormalization functions used are
given in Appendix~\ref{sec:appendix_renorm}; here we discuss their physical
motivation and qualitative features.

\begin{figure}[ht]
\centering
\includegraphics[width=\linewidth,keepaspectratio]{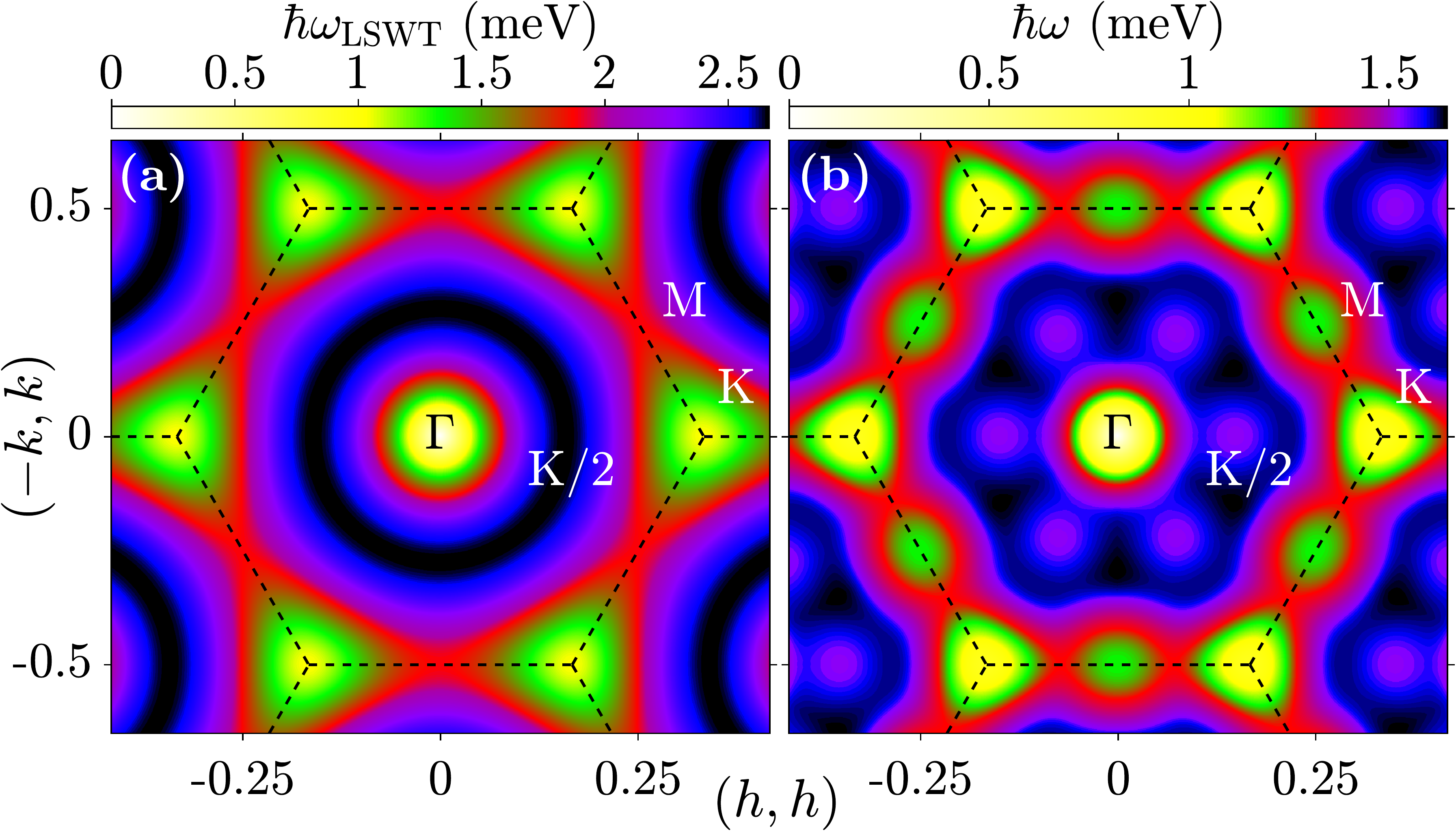}
\caption{(Color online) (a) Contour map in the $(hk0)$ plane of (a) the LSWT
dispersion $\hbar\omega_\mathrm{LSWT}(\vect{k})$ and (b) the best-fit
renormalized dispersion $\hbar\omega(\vect{k})$, using parameters in
Table~\ref{tab:pars} in Appendix~\ref{sec:appendix_renorm}. Dashed lines show
the hexagonal Brillouin zone boundaries. The separate color maps in the two
panels highlight relevant features of the two distinct dispersion surfaces. The
maximum in $\hbar\omega_\mathrm{LSWT}(\vect{k})$ occurs on a circle centered at
$\Gamma$ and passing near the six K/2 wave vectors; this is replaced in the
renormalized dispersion by a near-plateau in a wide annular region with shallow
local minima near the set of six K/2 points.
$\hbar\omega_\mathrm{LSWT}(\vect{k})$ has saddle points at the M zone-boundary
points, whereas at those positions the renormalized dispersion $\hbar\omega$ has
oval-shaped local minima. The triangular-shaped contours around K and
oval-shaped ones around M in (b) are clearly visible in the constant-energy INS
intensity map in Fig.~\ref{fig:hkslices}(g).
\label{fig:map_dispersion}}
\end{figure}

Wave vector dependent modifications are introduced to reproduce the local minima
(soft modes) observed in Fig.~\ref{fig:KMK_3meV_7meV} near the M$_2'$ point. The
lower soft mode occurs in the primary magnon dispersion. LSWT predicts a saddle
point at this position with a local maximum in the M-K direction and a local
minimum in the M-$\Gamma$ direction, see Fig.~\ref{fig:map_dispersion}(a). In
order to obtain a local minimum in both in-plane directions, we consider in
Fig.~\ref{fig:renorm} the mixing of the bare dispersion
$\hbar\omega_\mathrm{LSWT}$ (dashed magenta line) with a fictitious gapped mode
$\hbar\Omega_\mathrm{M}$ (dashed blue line) centered at M and parabolic in the
$hk$ plane; the resulting lower mode after mixing (magenta solid line) has the
desired qualitative feature of a smooth local minimum at M. To parametrize the
upper soft mode visible in Fig.~\ref{fig:KMK_3meV_7meV} near M$_2'$, we first
note that this feature occurs in the secondary modes $\omega^{\pm}(\vect{k})$,
which nearly overlap in this wave vector region and furthermore trade intensity
with each other, such that effectively a single higher-energy magnon branch is
visible. The corresponding location in reciprocal space where the primary mode
would display such a soft mode is near $\vect{k}_\mathrm{M}\pm\vect{Q}$,
symmetry equivalent to $\vect{Q}/2$, \textit{i.e.} located half-way between
$\Gamma$ and K; we will refer to this as K/2 from now on (in the notation of the
theoretical references \onlinecite{Verresen2019} and \onlinecite{Ferrari2019},
this is the Y$_1$ point). We illustrate in Fig.~\ref{fig:renorm} the procedure
to obtain a local soft minimum via mixing with a virtual parabolic mode
$\hbar\Omega_\mathrm{K/2}$ centered near K/2 (dashed green line); the lower mode
after mixing (magenta solid line) displays the desired local soft mode feature.
To obtain the ``final'' renormalized dispersion $\hbar\omega$ that was fitted to
the data, $\hbar\omega_\mathrm{LSWT}$ was mixed with many virtual paraboloids at
equivalent M and K/2-type positions (up to reciprocal lattice translations or
lattice point group symmetry operations) at the same $l$ value, in order to
ensure the final result is a smooth function that still respects all lattice
point group symmetries. Figure~\ref{fig:map_dispersion}(b) shows a contour map
of the renormalized dispersion surface in the $(hk0)$ plane, which highlights
the location of soft modes at M and near K/2 points, not present for the bare
$\hbar\omega_\mathrm{LSWT}$ dispersion in panel (a).

\begin{figure*}
\centering
\includegraphics[width=\linewidth,keepaspectratio]{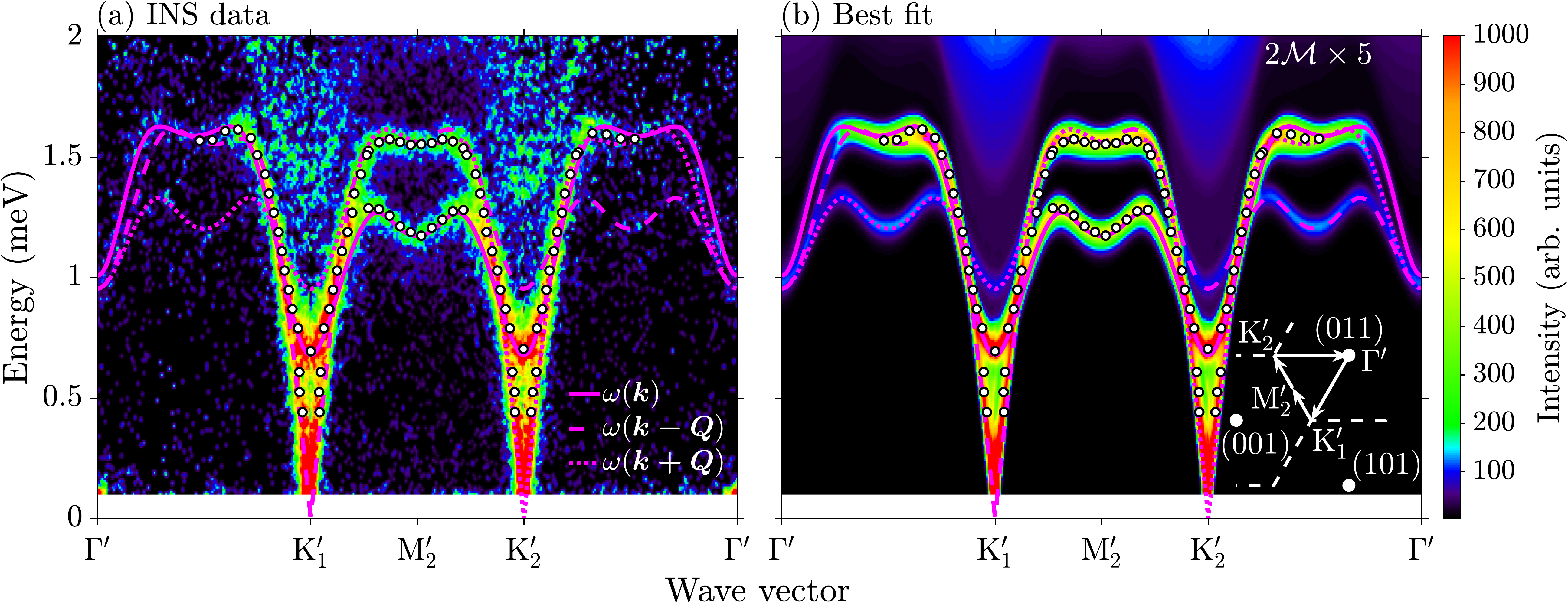}
\caption{(Color online) Comparison of (a) INS data (1.7~K,
$E_\mathrm{i}=3.53$~meV) with (b) the best-fit spin wave model with renormalized
dispersions, along a high-symmetry path in the $(hk1)$ plane crossing two
magnetic Bragg peaks at K$_{1,2}'$. The color in (a) shows the raw neutron
counts in arbitrary units. The calculation in (b) includes both the one-magnon
and two-magnon ($2{\mathcal{M}}$) cross-sections (the latter multiplied by a
factor of 5 for better visibility). The white circles in (a) are experimental
dispersion points (error bars smaller than the size of symbols), extracted by
fitting Gaussian peaks to constant energy or constant wave vector scans through
the sharp modes; their positions are well described by the model dispersions,
shown by the solid, dashed and dotted magenta lines for the $\omega(\vect{k})$,
$\omega(\vect{k-Q})$ and $\omega(\vect{k+Q})$ modes, respectively. The wave
vector path is $\Gamma' \, (0,1,1) \rightarrow \mathrm{K}'_1 \,
(\third{1},\third{1},1) \rightarrow \mathrm{M}'_2 \, (0,\half{1},1) \rightarrow
\mathrm{K}'_2 \, (-\third{1},\third{2},1) \rightarrow \Gamma'$, illustrated by
thick arrows in the white bottom-right inset with the Brillouin zone boundaries
represented by dashed lines. The integration width for the data in the $hk$
plane is $\pm0.05$~\AA$^{-1}$ and along $l$ is $\pm0.05$~\AA$^{-1}$. The
quasielastic scattering below 0.1~meV has been omitted.
\label{fig:KMK_slices}}
\end{figure*}

\begin{figure*}
\centering
\includegraphics[width=\linewidth,keepaspectratio]{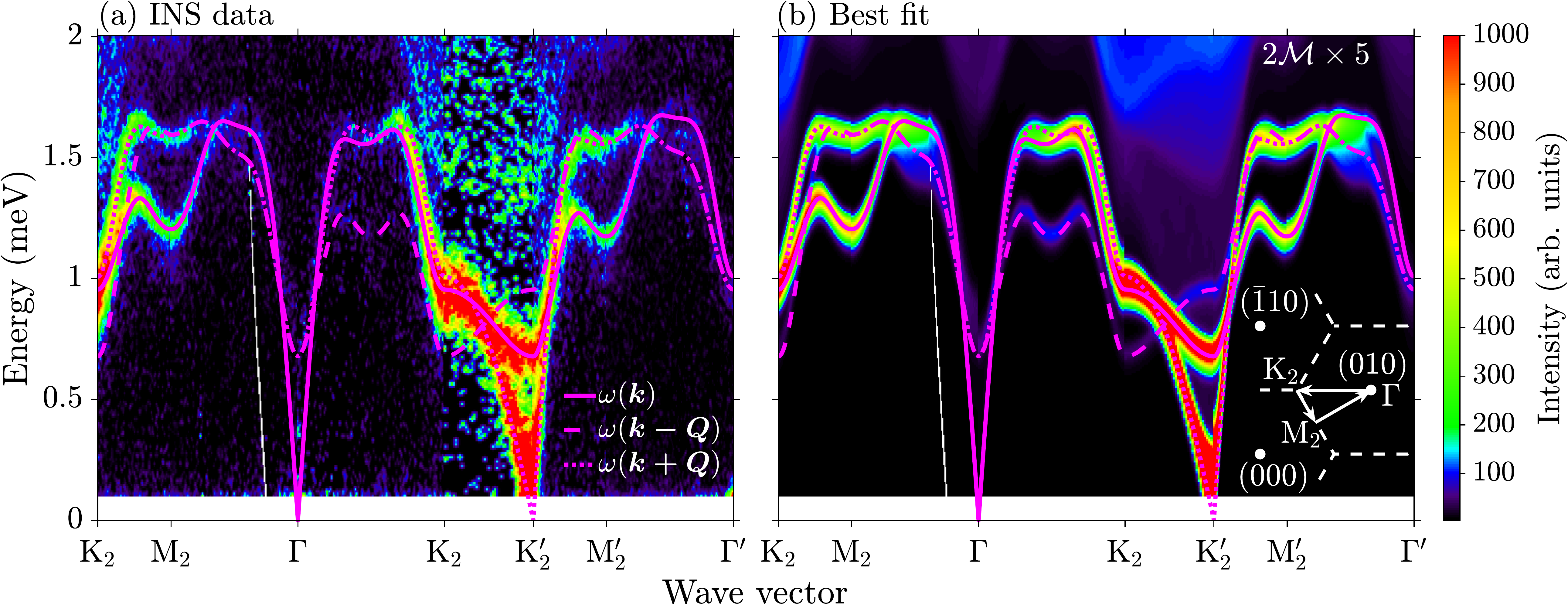}
\caption{(Color online) Same as Fig.~\ref{fig:KMK_slices}, but for a wave vector
path that probes the interlayer $l$ dispersion: $\mathrm{K}_2 \,
(-\third{1},\third{2},0) \rightarrow \mathrm{M}_2 \, (0,\half{1},0) \rightarrow
\Gamma \, (0,1,0) \rightarrow \mathrm{K}_2 \rightarrow \mathrm{K}'_2 \,
(-\third{1},\third{2},1) \rightarrow \mathrm{M}'_2 \, (0,\half{1},1) \rightarrow
\Gamma' \, (0,1,1)$, illustrated in the white bottom-right inset. The
integration width in the $hk$ plane is $\pm0.05$~\AA$^{-1}$ for panels 1--3 and
6, and $\pm0.02$~\AA$^{-1}$ for panels 4 and 5, and along $l$ is
$\pm0.1$~\AA$^{-1}$ for panels 1, 3 and 5, and $\pm0.3$~\AA$^{-1}$ for panels 2
and 6.
\label{fig:KMG_slices}}
\end{figure*}

\subsection{Fits of INS data to the spin wave model with renormalized
dispersions}
\label{sec:fits}

The above spin wave model with renormalized dispersions was fitted to the
experimental data as follows. First, an initial parametrization of the
dispersion relation was obtained by fitting the functional form of the
renormalized spin wave dispersion to a set of $(h,k,l,E)$ dispersion points,
extracted by fitting Gaussian peaks to constant energy or constant wave vector
scans through the INS data in regions where the magnon modes were clearly
separated from one another and where the character of each mode [whether
$\omega(\vect{k})$, $\omega^{+}(\vect{k})$ or $\omega^-(\vect{k})$] could be
unambiguously identified from the dispersion trends. This dispersion
parametrization was then used as a starting point and further refined by
performing a global fit of the full one-magnon cross-section model, including
all three magnon branches, to selected slices and cuts through the
four-dimensional INS data along many symmetry-distinct directions in reciprocal
space (representative slices shown in Figs.~\ref{fig:KMK_slices} and
\ref{fig:KMG_slices}). To ensure the model fitted only the one-magnon intensity
data, the regions with clear continuum scattering in those slices were masked in
the fit; for example, data pixels contributing to the gapped ``cones'' of
continuum scattering at high energies near K$_{1,2}$ points in
Fig.~\ref{fig:KMK_slices}(a) were excluded from the fit. The one-magnon
cross-section model included the effects of the finite-temperature Bose factor,
the magnetic form factor for Co$^{2+}$ ions, the neutron polarization factor,
and a parametrization of the exprimental energy resolution (for details, see
Appendix~\ref{sec:appendix_LSWT}). The linewidth of the observed sharp
one-magnon modes in constant wave vector scans was well accounted for by the
parametrized instrumental energy resolution, suggesting that the magnons are
long-lived with no evidence of lifetime broadening. Model parameters obtained
through this fitting procedure are listed in Table~\ref{tab:pars}
(Appendix~\ref{sec:appendix_renorm}) and include the two exchange parameters
$J_1$ and $J_z$, the exchange anisotropy $\Delta$, the relative
in-plane/out-of-plane magnon intensity prefactor $Z_{\eta}/Z_{\xi}$, and
parameters to describe the two types of soft modes at M and near K/2. The
Hamiltonian parameters were constrained to reproduce the observed magnetization
saturation field\cite{Kamiya2018}
\begin{equation}
g\mu_{\rm{B}}B_{\rm sat}S^{-1}=(3+6\Delta)J_1+2(1+\Delta)J_z,
\label{eq:constraint}
\end{equation}
with $B_{\rm sat}=32.8$~T and $g$ factor $g=3.87$.

This model provides an excellent description of the experimental dispersion
relations at all energies and wave vectors probed, as illustrated by comparing
(a) the data and (b) the parametrization plots in Fig.~\ref{fig:KMK_slices} for
wave vector directions in the $(hk1)$ plane and Fig.~\ref{fig:KMG_slices} for
wave vectors also probing the interlayer $l$ dispersions. Open white circles in
Fig.~\ref{fig:KMK_slices}(a) correspond to empirical peak centers extracted from
Gaussian fits to constant energy or constant wave vector scans; their close
agreement with the overplotted dispersion relations (magenta lines) emphasises
the level of quantitative agreement between data and model. All key features of
the dispersion are quantitatively reproduced, including the energy of the gapped
mode at the magnetic Bragg peak positions K$_{1,2}'$, the dispersion along the
interlayer  K$_2$-K$'_2$ direction in Fig.~\ref{fig:KMG_slices}, the relative
flattening of the dispersions near the maximum energy, and the dispersion shapes
near the soft modes at M and near K/2 points.

Although the present analysis focuses on capturing the intermediate to
high-energy features of the dispersions, where the spin wave peaks are most
accurately determined experimentally as they are well separated in energy and
momentum, the parametrization also captures well the low-energy behavior. In
particular, the steep linearly-dispersive spin wave cones emerging out of the
magnetic Bragg peak positions K$_{1}'$ and K$_{2}'$ in
Fig.~\ref{fig:KMK_slices}, attributed to the gapless $\omega^{-}(\vect{k})$ and
$\omega^{+}(\vect{k})$ modes, respectively, are consistent between the data and
the model parametrization. We note however that the spin wave peaks are barely
resolved at low energies due to the very steep dispersion combined with the
finite instrumental resolution, so changes in the spin wave velocity of order
$10\%$ compared to the LSWT result as predicted by SWT$+1/S$
treatments\cite{Chubukov1994} could also be consistent with the data in this
low-energy region. Testing quantitatively for such spin wave renormalization
effects in the limit $\omega \rightarrow 0$ would require a more sophisticated
analysis, including theoretical predictions of the complete wave vector and
energy-dependent quantum renormalization of the dispersions and intensities for
the full Hamiltonian in Eq.~(\ref{eq:Hamiltonian}), which is beyond the scope of
the present empirical parametrization.

Turning now to the magnon intensities, the strongest signal in
Figs.~\ref{fig:KMK_slices} and \ref{fig:KMG_slices} is observed near K points
with intensities decreasing rapidly approaching the $\Gamma$ points, and this
general trend is well reproduced by the model. However, close inspection of the
intensity variation, in particular as a function of energy, reveals a
discrepancy between the data and model; namely, if the overall intensity scale
in the calculation is set to match the intensities of the low-energy magnons in
those figures, then the intensity of the high-energy magnons is much lower in
the data than in the calculation, compare Fig.~\ref{fig:KMK_slices}(a) with (b),
also Fig.~\ref{fig:KMG_slices}(a) with (b), and Fig.~\ref{fig:hkslices}(h) with
(d). (Unless otherwise specified, the overall intensity scale is chosen to match
the observed low-energy signal for all calculated intensity color maps
throughout the paper.) We propose that this discrepancy between the spin wave
model and data is evidence of a transfer of spectral weight from the one-magnon
modes to the higher-energy continuum scattering that is not captured by the
model; such a transfer of spectral weight is expected from general
considerations as a consequence of the interaction between the high-energy
magnons and the higher-energy continuum states, expected to result in a downward
renormalization of the magnon energies and a simultaneous transfer of intensity
from the high-energy magnons to the continuum states. Further support for this
interpretation will be provided later in Section~\ref{sec:two_magnon}, where we
compare directly the observed scattering lineshapes with predictions of the spin
wave model for both one- and two-magnon excitations.

\begin{figure*}
\centering
\includegraphics[width=0.8\linewidth,keepaspectratio]{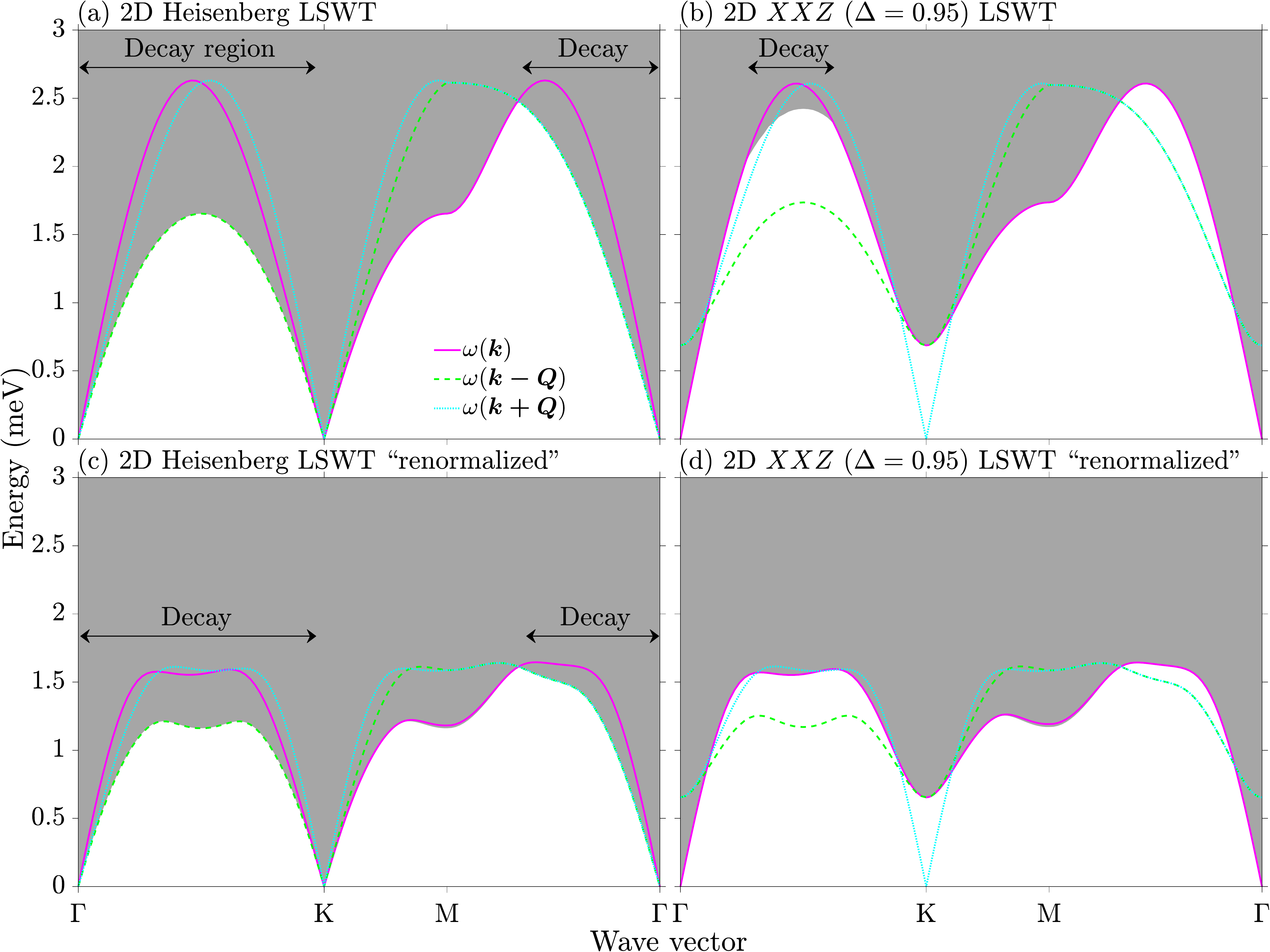}
\caption{(Color online) Phase space (shaded area) for two-magnon excitations
along high-symmetry wave vector directions, compared with the one-magnon
dispersion relation $\hbar\omega(\vect{k})$ (magenta solid line). When overlap
occurs, magnons are unstable to decay, and those regions are indicated by the
horizontal double-arrowed lines labeled ``Decay''. The four figure panels
correspond to different two-dimensional spin wave models related to the
Hamiltonian in Eq.~(\ref{eq:Hamiltonian}) with $J_z=0$. (a) 2D Heisenberg
($\Delta=1$) model within LSWT, for which decay is expected along the whole
$\Gamma$-K line and also a finite wave vector range starting from $\Gamma$
towards M. (b) Easy-plane XXZ ($\Delta=0.95$) model within LSWT, where the
anisotropy opens a gap at K. The phase space for magnon decay is much reduced
compared to (a), but is still present near the top of the $\Gamma$-K dispersion.
(c) and (d) are the same as (a) and (b), but with the empirical renormalizations
applied to the LSWT dispersions, as described in Sec.~\ref{sec:renorm}. In (c)
the renormalizations have not greatly affected the decay regions compared to
(a), whereas in (d) the decay regions are completely eliminated, meaning magnons
are sharp everywhere. The exchange and dispersion renormalization parameters
used are listed in Table~\ref{tab:pars} in Appendix~\ref{sec:appendix_renorm}.
As explained in the text, the plots are for the {\em rotating} reference frame,
where a single magnon mode is present with dispersion $\hbar\omega(\vect{k})$.
\label{fig:lswt_two_magnon}}
\end{figure*}

\subsection{Why are magnons sharp and do not decay?}
\label{sec:one2two}

We find experimentally that the magnons are sharp, with resolution-limited
lineshapes throughout the extensive region of reciprocal space probed with no
evidence of intrinsic broadening, indicating that magnon decay processes do not
occur. This is a non-trivial result, as SWT+$1/S$ theoretical studies have
predicted extended regions of one$\rightarrow$two magnon decays for the
spin-$\half{1}$ TLHAF limit.\cite{Mourigal2013} We review below the requirements
for magnon decays following Ref.~\onlinecite{Chernyshev2009} and find that they
are not satisfied in \BCSO{}. In particular, we find that the shape of the
magnon dispersion is quite different from that of the TLHAF model and is such
that overlap between one- and two-magnon phase spaces is avoided throughout
reciprocal space, so no decay can occur.

Specifically, decay processes require that (i) the spin Hamiltonian has finite
matrix elements for mixing between one- and two-magnon states, and (ii) energy
and momentum are conserved during the decay, \textit{i.e.} a magnon at wave
vector $\vect{k}$ can kinematically decay into a pair of magnons with wave
vectors $\vect{k}_1$ and $\vect{k}_2$ if
\begin{equation}
\vect{k}=\vect{k}_1+\vect{k}_2 \quad \mathrm{and} ~~
\omega(\vect{k})=\omega(\vect{k}_1)+\omega(\vect{k}_2)\nonumber.
\end{equation}
The finite matrix element requirement for decay is naturally satisfied due to
the non-collinear nature of the $120\ang$ order in the ground state, which leads
to couplings between longitudinal spin fluctuations on one site and transverse
fluctuations on neighboring sites (defining longitudinal and transverse as along
and perpendicular to the local ordered spin direction, respectively), which in
turn mixes one- and two-magnon states.\cite{Chernyshev2009} The kinematic
constraint is most transparently tested by working in the \textit{rotating}
reference frame that follows the local ordered spin orientation, in which the
ground state is ferromagnetic and there is a single magnon mode with dispersion
$\omega({\vect{k}})$ (for more details see Appendix~\ref{sec:appendix_LSWT}).
The phase space of two-magnon excitations in this rotating frame is illustrated
by the shaded area in Fig.~\ref{fig:lswt_two_magnon}(a) for the 2D Heisenberg
model described within LSWT. Decay is expected where the one-magnon dispersion
(magenta solid line) overlaps with the shaded area, which occurs throughout the
$\Gamma$-K line and over a significant portion of the $\Gamma$-M line. In
addition to the magnon dispersion $\omega({\vect{k}})$, the figure also shows
the wave vector shifted dispersions $\omega({\vect{k}\pm\vect{Q}})$ (dotted
green/dashed cyan lines), which helps to emphasize that the lower boundary of
the two-magnon continuum at a fixed wave vector $\vect{k}$ is the minimum energy
of those three curves. This occurs because the lower boundary corresponds to
creating one of the two magnons at zero energy at either the $\Gamma$ point
($\vect{k}_1=\vect{0}$) or at one of the two K points
($\vect{k}_1=\mp\vect{Q}$), thus placing the other magnon in the pair at wave
vector $\vect{k}_2=\vect{k}$ with energy $\omega(\vect{k})$ or at
$\vect{k}_2=\vect{k}\pm\vect{Q}$ with energy $\omega(\vect{k}\pm\vect{Q})$,
respectively. If an easy-plane anisotropy is added, as in
Fig.~\ref{fig:lswt_two_magnon}(b), the dispersion becomes gapped at the ordering
wave vector (K points), which increases the minimum energy cost of creating
two-magnon states and therefore reduces the regions of overlap between one- and
two-magnon states. Despite this, a finite decay region is still expected near
the top of the $\Gamma$-K dispersion, if the dispersion shape is given by the
LSWT result.

\begin{figure}
\centering
\includegraphics[width=\linewidth,keepaspectratio]{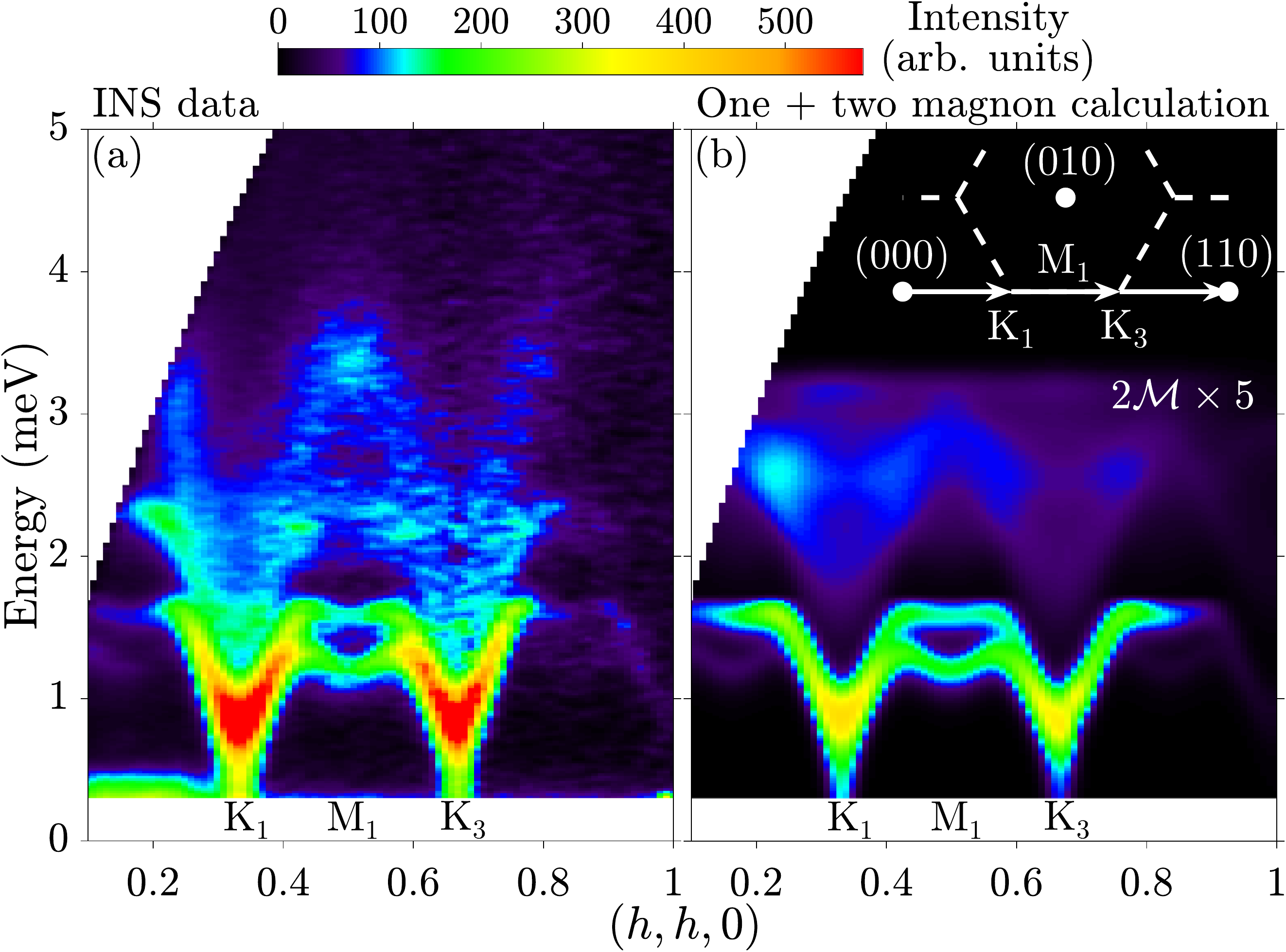}
\caption{(Color online) (a) INS data (1.7~K, $E_\mathrm{i}=7.01$~meV) showing
the full extent of the magnetic excitation spectrum, compared in (b) with the
renormalized spin wave model. The wave vector path is
$(h\mp0.05,h\pm0.05,\pm2.39)$, where the $\pm$ signs indicate the data
integration range. The color shows the raw neutron counts in arbitrary units.
The calculation in (b) includes both the one-magnon and two-magnon
($2{\mathcal{M}}$) cross-sections (the latter multiplied by a factor of 5 for
better visibility). The intensity scale factor in (b) is chosen to reproduce the
observed intensities of the sharp magnons near M$_1$ in (a).
\label{fig:110_slices_7meV}}
\end{figure}

The above analysis is however oversimplified, as the experimental dispersion
relations are in fact strongly renormalized in non-trivial ways compared to the
LSWT prediction, as found in the preceding Sec.~\ref{sec:fits}. This is
physically attributed to the effect of magnon interactions and quantum
fluctuations beyond the linear spin wave approximation. In
Fig.~\ref{fig:lswt_two_magnon}(c), the solid magenta line is the dispersion
relation from (a) after applying the same wave vector-dependent renormalizations
as for the full model fitted to the experimental data, using the parameters in
Table~\ref{tab:pars} but with $J_z=0$ and $\Delta=1$. In other words, we assume
that the empirically-determined dispersion renormalizations are unaffected by
the weak 3D couplings and the small easy-plane anisotropy.
Figure~\ref{fig:lswt_two_magnon}(c) shows that the overlap regions are not
changed much by these renormalizations and so extended decay regions are
predicted. Finally, in Fig.~\ref{fig:lswt_two_magnon}(d) we consider a spin wave
model with finite easy-plane anisotropy and dispersion renormalizations
included, which is closer to experimental observations. In this case, we find
that the magnon dispersion $\omega(\vect{k})$ defines the lower boundary of the
two-magnon continuum but never enters it, so decay regions are completely
eliminated. For finite values of the interlayer coupling $J_z$ that are
consistent with the experimental data, there are only very small changes to
cases (c) and (d) and the qualitative content is unaffected, \textit{i.e.}
extended decay regions are still present in (c) but remain absent in (d). As the
3D couplings have only very small effects on the magnon decay regions, panel (d)
captures the essential physics of avoided magnon decays in the present system.

The above analysis of the different models suggests that magnon decays do not
occur in \BCSO{} because of the combined effect of the small, but finite
easy-plane anisotropy ($\Delta<1$) and the strong dispersion renormalizations
from quantum effects, with both effects playing a r\^{o}le.

We note that recent theoretical work,\cite{Verresen2019} based on DMRG
calculations for gapped spin models models slightly perturbed away from the
TLHAF limit, proposed that strong quantum interactions lead to an avoidance of
the LSWT-predicted overlap between the one-magnon dispersion and the
higher-energy two-magnon continuum scattering; the resulting magnon dispersion
is renormalized downwards and has a much reduced spectral weight, due to a
transfer of weight to the higher-energy continuum states via the aforementioned
interactions. It would be interesting if such calculations could be extended to
the weak easy-plane anisotropy case relevant here where the spectrum is gapless,
and also much closer to the isotropic Heisenberg limit, to test if the same
picture applies. In addition, recent variational dynamical Monte Carlo
calculations proposed that magnons remain sharp throughout the Brillouin zone in
the fully isotropic Heisenberg limit.\cite{Ferrari2019}

\subsection{Continuum scattering compared with a two-magnon cross-section}
\label{sec:two_magnon}

An overview of the complete magnetic excitation spectrum is plotted in
Fig.~\ref{fig:110_slices_7meV}(a). Sharp spin wave modes are visible up to
1.6~meV, followed by a continuum of scattering, which appears to emerge from
inside the spin wave cones centered at wave vectors K$_{1,3}$ and extends in
energy up to at least the top of the plotted range. Inside the continuum,
highly-dispersive intensity modulations are clearly visible, in the form of two
successive cones of intensity in different energy ranges, both centered at the K
points and dispersing in energy with maxima at M points. Panel (b) shows the
corresponding calculation for the best-fit renormalized spin wave model. The
magnon dispersions are well captured, but the predicted two-magnon
(2$\mathcal{M}$) continuum (shown with intensity scaled up by a factor of 5 for
visibility) is not able to account for the large scattering weight in the
experimentally observed continuum. Nor can it explain the highly-structured
intensity modulations, predicting just one filled cone of intensity centered at
K points and dispersing in energy up to M, shifted in energy compared to the
experimentally-observed intensity modulations in panel (a).

\begin{figure}
\centering
\includegraphics[width=\linewidth,keepaspectratio]{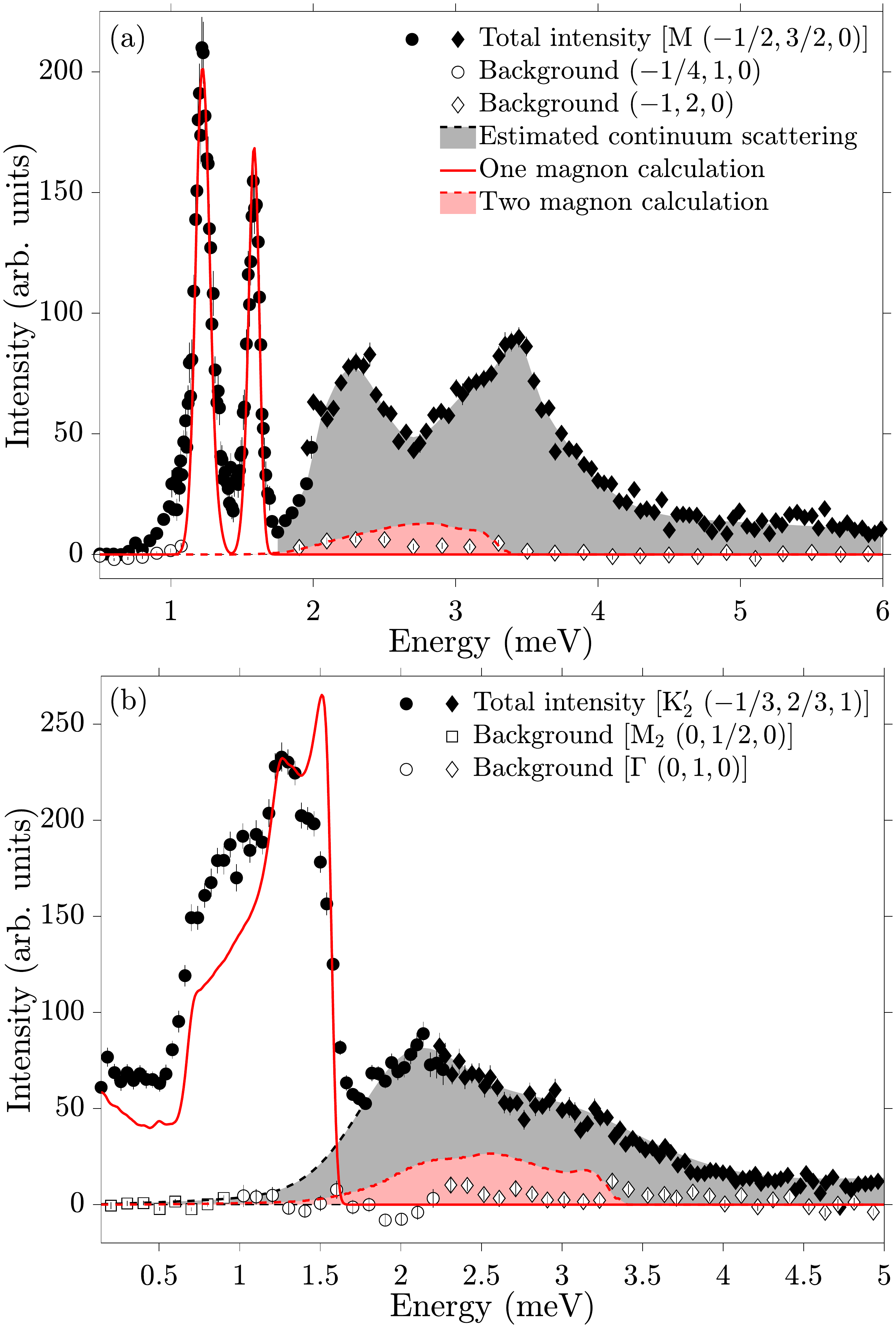}
\caption{(Color online)
Energy scans through the INS data (filled symbols, 1.7~K) (a) at an M zone
boundary point [equivalent to M$_1$ in Fig.~\ref{fig:110_slices_7meV}(a)] and
(b) near a magnetic Bragg wave vector (K$_2'$ in Fig.~\ref{fig:KMK_3meV_7meV}),
compared with the renormalized spin wave model (red line and pink shading
represent one-magnon and two-magnon excitations, respectively). Data points are
raw neutron counts with an estimate of an energy-dependent non-magnetic
background subtracted. The accuracy of the background subtraction is illustrated
by the open symbols, which show the resulting intensities in regions of wave
vector and energy where no magnetic scattering is expected. Circles (diamonds)
correspond to $E_\mathrm{i}=3.53$~meV (7.01~meV) data. The wave vector
integration range extends over the full available $l$ range (several zones), and
in the $hk$ plane is centered at the nominal wave vector; the range in (a) is
$\pm0.075$ along both $(1,0,0)$ and $(-\half{1},1,0)$, and in (b) it is a
circular region illustrated in Fig.~\ref{fig:KMK_3meV_7meV} (lower-left inset).
\label{fig:en_scans}}
\end{figure}

Figure~\ref{fig:en_scans}(a) presents a quantitative data vs. model lineshape
comparison for an energy scan at a wave vector equivalent to M$_1$ [near the
center of Fig.~\ref{fig:110_slices_7meV}(a)]. The two sharp peaks on the
low-energy side are well accounted for by resolution-limited magnons, where the
first peak is identified with the out-of-plane $\omega({\vect{k}})$ mode and the
second with degenerate in-plane $\omega^{\pm}(\vect{k})$ magnons. However, the
large continuum scattering at higher energies (emphasized by the gray shading)
is much underestimated by the two-magnon cross-section (pink shading). (For
details of the calculation, see Appendix~\ref{sec:appendix_LSWT}.) Note that the
two prominent broad peaks in the continuum near 2.3 and 3.5~meV correspond to
the two broad intensity maxima near the center of
Fig.~\ref{fig:110_slices_7meV}(a).

Another useful comparison is provided in Fig.~\ref{fig:en_scans}(b) by an energy
scan at a magnetic Bragg peak position (K$_2'$ in Fig.~\ref{fig:KMK_3meV_7meV}).
Key features of the one-magnon spectrum are well reproduced (red line), such as
the flat signal at the lowest energies, due to the gapless $\omega^+(\vect{k})$
mode, and the rapid intensity increase near 0.7~meV, due to intersecting the
gapped $\omega(\vect{k})$ mode. However, the relative intensity between high-
and low-energy magnons is overestimated, \textit{i.e.} if the intensity scale
were set to match the signal below 0.7~meV in Fig.~\ref{fig:en_scans}(b) then
the high-energy magnons would have been greatly overestimated; we interpret this
as evidence for a transfer of spectral weight from the high-energy magnons to
the higher-energy continuum scattering, not captured by the spin wave model. The
gray shading highlights the continuum scattering contribution, which is much
underestimated by the two-magnon calculation (pink shading with dashed line
envelope). We propose that the enhanced scattering continuum is at least
partially due to the transfer of spectral weight from the high-energy magnons.

\begin{figure}
\centering
\includegraphics[width=\linewidth,keepaspectratio]{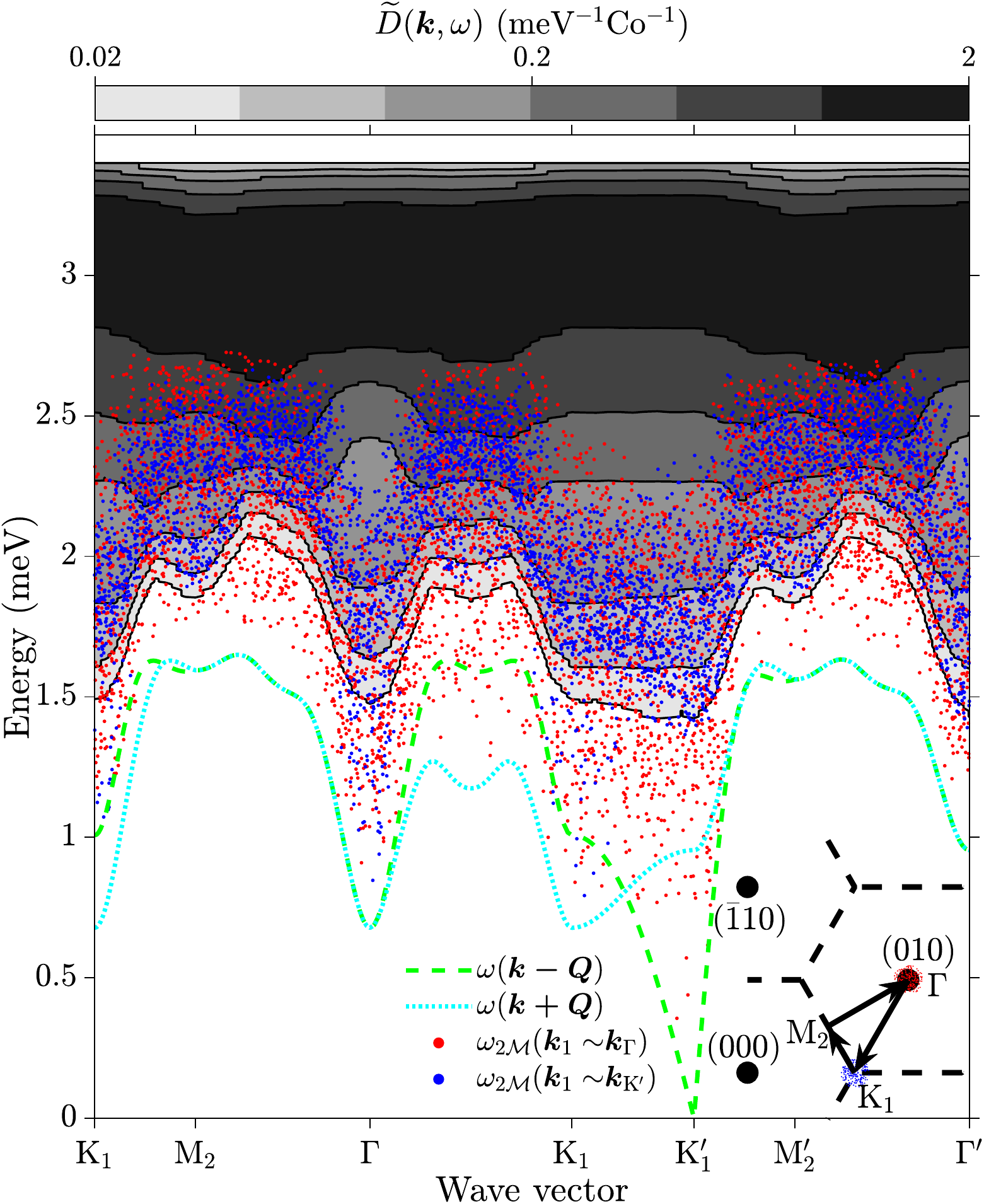}
\caption{(Color online) Gray shaded contour plot ($\log_{10}$ scale) of the
two-magnon density of states in Eq.~(\ref{eq:DoS}) along a wave vector path
equivalent to the one in Fig.~\ref{fig:KMG_slices}(b). The density of states is
very small (below the plotted gray range) in the region immediately above the
lower continuum boundary, given by the lower energy of the curves
$\omega^-(\vect{k})$ and $\omega^+(\vect{k})$, plotted by dashed green and
dotted cyan lines, respectively. The dominant two-magnon states that contribute
in this region have one of the magnons near the origin, with a very low density
of states in energy (sparsely-distributed red dots). At higher energy, more
two-magnon scattering channels become available, such as having one magnon near
K (blue dots), leading to a significant increase in the density of states above
the lowest black contour line. The colored dotted regions near $\Gamma$ and
K$_1$ in the bottom-right diagram indicate the phase spaces sampled by the
two-magnon events plotted as dots in the main panel.
\label{fig:KMG_slices_DoS}}
\end{figure}

\begin{figure*}
\centering
\includegraphics[width=0.9\linewidth,keepaspectratio]{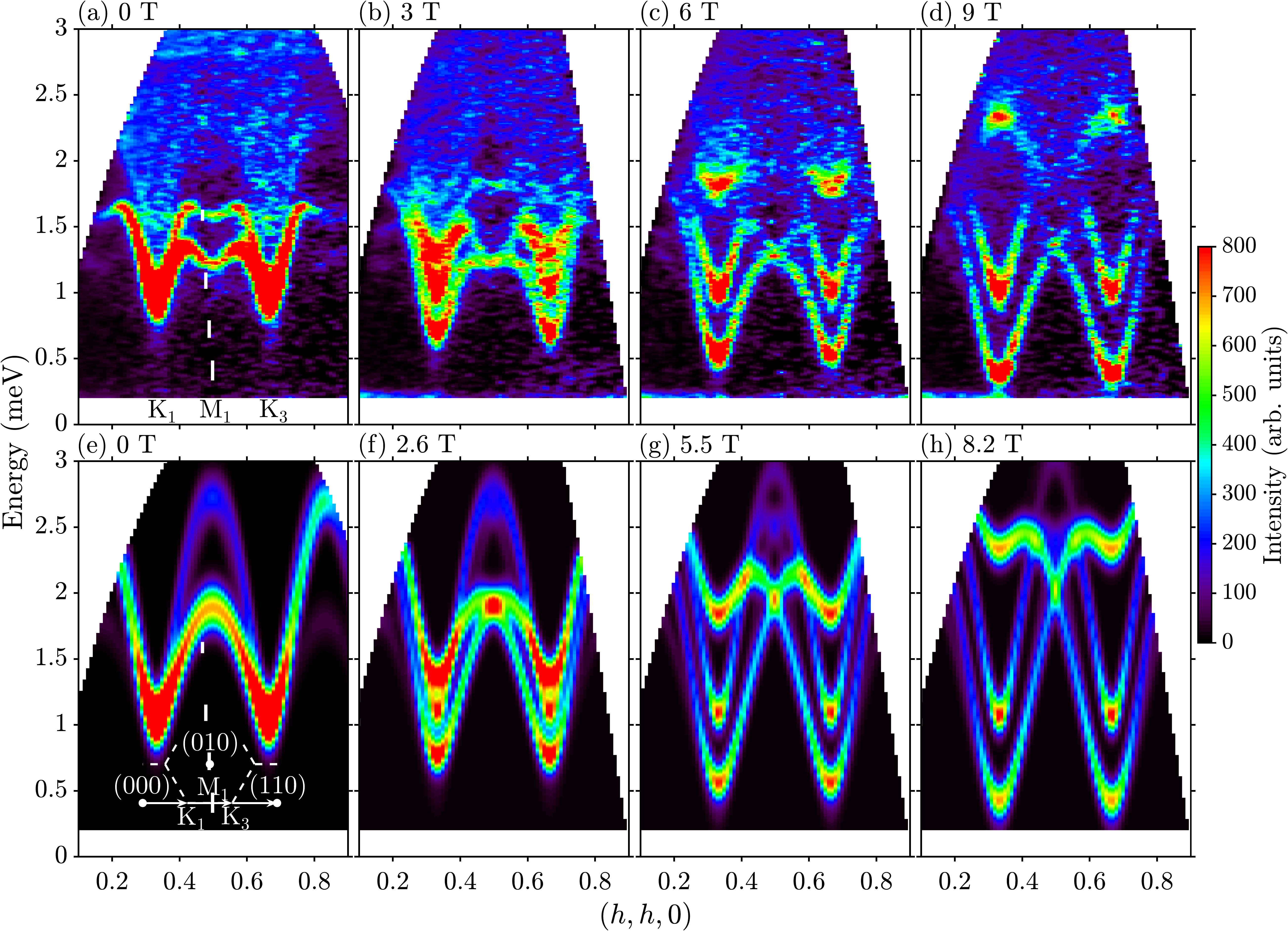}
\caption{(Color online) INS data (top row) as a function of $c$-axis applied
magnetic field, compared with the predicted spin wave spectrum (bottom row) for
a sample containing equal-weight magnetic domains of types 1 and 2 in
Fig.~\ref{fig:Structure}(c). The color is the intensity in arbitrary units and
the wave vector path is $(h\mp0.05,h\pm0.05,\pm0.3)$, where $\pm$ values
indicate the integration ranges. Panel (a) was collected with
$E_\mathrm{i}=3.53$~meV at 1.7~K and (b)--(d) with $E_\mathrm{i}=3.81$~meV at
0.1~K. (e)--(h) Corresponding spin wave spectra within LSWT using the
Hamiltonian parameters\cite{Ito2017} $J_1=1.67$~meV, $J_z=0.08$~meV, $g=4$ and
$\Delta=0.954$, assuming no quantum renormalizations of the dispersions. For
each panel, the intensity scale and magnetic field value (listed in the panel
titles) were selected to give the best agreement with the data for energy scans
at K$_1$. The quasielastic scattering below 0.2~meV has been omitted.
\label{fig:110_slices_infield}}
\end{figure*}

Close inspection of Fig.~\ref{fig:KMK_3meV_7meV} shows that the continuum of
scattering appears to be separated in energy from the one-magnon modes at lower
energies. We propose below that the most likely explanation of this effect is a
suppression of the density of states for two-particle continuum scattering,
rather than a genuine energy gap between the two types of excitations. The
energy separation is most apparent in the center of the figure at M$_2'$, where
the highest-energy magnon is at 1.65~meV, whereas significant continuum
scattering starts only above about 1.8~meV. This separation is reduced (but
still present) inside the spin wave cones centered at K$_{1,2}'$, as significant
continuum scattering does not start immediately above the sharp modes and there
is a clear drop in intensity between the two scattering signals. Note that an
energy separation between the two-magnon continuum and the magnon modes is also
clearly visible in the spin wave model calculation in
Figs.~\ref{fig:KMG_slices}(b) and \ref{fig:110_slices_7meV}(b). This apparent
separation in the calculation seems to be at odds with the fact that the magnon
spectrum is gapless [as there is a Goldstone mode at the $\Gamma$ point
$\omega(\vect{0})=0$ associated with rotations of the ordered spins in the $ab$
plane], so one can always create a magnon pair excitation at the wave vector and
energy of a single magnon (by creating one magnon in the pair at the origin);
therefore, no energy gap is expected between one-magnon states and the
two-magnon continuum, as illustrated in Fig.~\ref{fig:lswt_two_magnon}(d).
Indeed, close inspection of energy scans through Figs.~\ref{fig:KMG_slices}(b)
and \ref{fig:110_slices_7meV}(b) shows that no finite gap is present, the
continuum intensity is just very small immediately above the one-magnon
dispersions. This is because the relevant two-magnon states that contribute just
above the lower boundary of the continuum are dominated by pairs where one
magnon is created close to zero energy near the origin; since the dispersion
there is very steep [see Fig.~\ref{fig:KMG_slices}(b) solid line near $\Gamma$],
the density of states in energy for such two-magnon processes is very small,
leading to an apparent suppression of the two-magnon signal near the lower
boundary. This is illustrated in Fig.~\ref{fig:KMG_slices_DoS}, where the gray
shadings separated by black lines in the top half of the graph illustrate a
contour map (on a log scale) of the two-magnon density of states
$\widetilde{D}(\vect{k},\omega)$ in Eq.~(\ref{eq:DoS}). Note that the region
immediately above the lower boundary onset (the lower of the dashed green and
dotted cyan lines) is below the plotted gray range, indicating a very low
density of states. The sparsely-distributed red dots correspond to two-magnon
states where one magnon is near the origin, showing that two-magnon continuum
states do exist just above the magnon dispersions. However, their density is
very low compared to higher energies, for example above the lowest black contour
line, where new scattering channels become available and there is a significant
contribution from pair states with one magnon near the (gapped) K point (blue
dots). Based on this analysis, we conclude that the apparent separation in the
data between the higher-energy continuum scattering and the lower-energy sharp
spin wave modes is consistent with the assumed gapless spin wave spectrum and is
most likely due to a suppression of intensity towards the lower boundary of the
continuum due to a reduced density of states in that region.

\begin{figure*}
\centering
\includegraphics[width=0.8\linewidth,keepaspectratio]{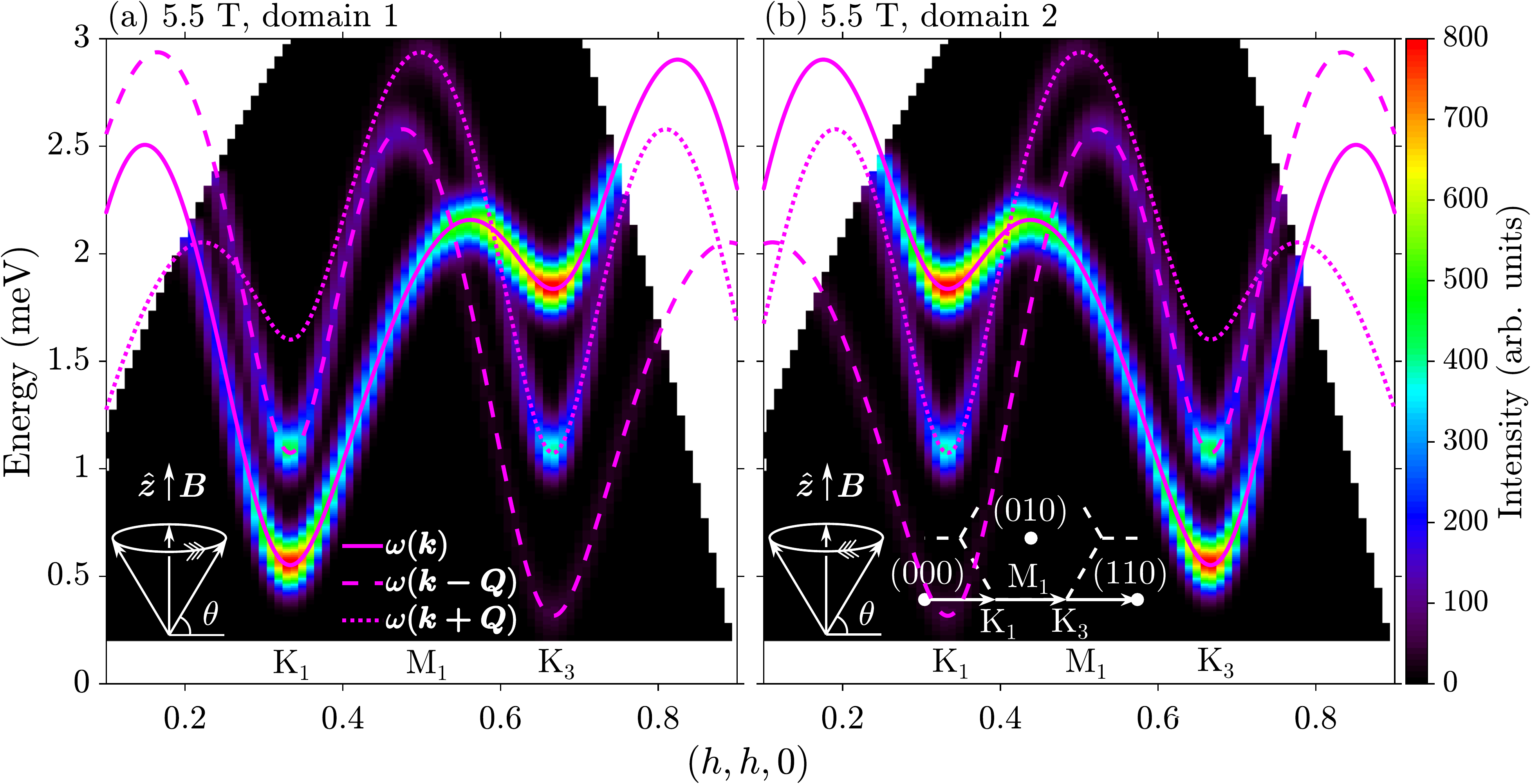}
\caption{(Color online) Spin wave spectrum in the cone phase in a $c$-axis
applied field for (a) domain 1 and (b) domain 2, to be compared with the data in
Fig.~\ref{fig:110_slices_infield}(c). The two domains have different spectra,
mirror-reversed about the zone boundary M$_1$ point. The color is the one-magnon
cross-section from Appendix~\ref{sec:appendix_LSWT}, for the same parameters as
in Fig.~\ref{fig:110_slices_infield} (g). The white bottom-left inset in each
panel shows the sense of rotation of the spins with respect to the applied field
$\vect{B}$ (vertical arrow), as per Fig.~\ref{fig:Structure}(c). The
$\omega(\vect{k})$, $\omega(\vect{k-Q})$ and $\omega(\vect{k+Q})$ dispersions
are plotted as solid, dashed and dotted magenta lines, respectively.
\label{fig:110_slices_chiral}}
\end{figure*}

\section{Magnon dispersions in the cone phase in c-axis magnetic field}
\label{sec:finite_field}

Here we present INS measurements of the magnetic excitations as a function of
magnetic field applied along the $c$ axis, which are sensitive to the presence
of multiple magnetic domains. For the Hamiltonian in Eq.~(\ref{eq:Hamiltonian}),
the mean-field ground state has ordered spins rotating by $120\ang$ between NN
sites in the triangular layers, with two possible senses of rotation illustrated
in Fig.~\ref{fig:Structure}(c) left/right panel, corresponding to a
counterclockwise/clockwise rotation between sites displaced along the
$\vect{a}+\vect{b}$ direction (labeled `domain 1'/`domain 2'), respectively. The
two structures are degenerate in the absence of DM interactions, so a
macroscopic sample would be expected to contain magnetic domains of both types,
selected via spontaneous symmetry breaking when cooling through the magnetic
ordering temperature. In zero magnetic field, the two domains have identical
dispersion relations and dynamical structure factors. In a $c$-axis applied
magnetic field, spins cant towards the field while their in-plane component
continues to rotate in the $ab$ plane, forming a cone structure. The two domains
remain degenerate in applied field, but their excitation spectrum is different,
as the primary magnon dispersion $\omega(\vect{k})$ acquires an additive term
[$C_{\vect{k}}$ in Eqs.~(\ref{eq:Ck}) and (\ref{eq:Dispersion}) in
Appendix~\ref{sec:appendix_LSWT}] that changes sign between the two domain
types. Previous magnetization,\cite{Susuki2013} nuclear magnetic
resonance\cite{Koutroulakis2015} and ultrasound velocity\cite{Quirion2015}
measurements in $c$-axis applied field have indicated that the cone phase
persists up to 12~T. Here we present measurements well within this field range
(up to 9~T) to test whether the sample contains both types of domains, selected
via spontaneous symmetry breaking, as expected in the absence of DM
interactions.

Figure~\ref{fig:110_slices_infield} (top row) shows how the magnetic excitations
along a representative wave vector path evolve upon increasing the applied
field. In zero field [panel (a)], the spectrum has mirror symmetry around the
zone boundary M$_1$ point with two intense, gapped spin wave modes visible,
clearly separated at M$_1$ and nearly overlapping near the K$_{1,3}$ points,
followed by continuum scattering at higher energies. In the following we focus
on the sharp spin wave modes, as they contain the key information about the
domain type. At 3~T [panel (b)], there are clearly three modes resolved near the
K points, which separate further upon increasing field to 6, then 9~T [panels
(c) and (d)], with the mirror symmetry of the spectrum around M$_1$ preserved
throughout. The data presented have contributions from pixels at wave vectors
$\vect{k}$ not only along the nominal $(110)$ scan direction, but also along
other directions in the $(hk0)$ plane that are equivalent up to symmetry
operations of the crystal lattice point group; this was performed for the
purpose of improving the counting statistics. We have explicitly verified that
slices through the raw, unsymmetrized data display all the same features.

In order to interpret the observed behavior, we compare in 
Fig.~\ref{fig:110_slices_chiral} the predicted spectrum within LSWT for magnetic
domains of both types at a representative intermediate field where the mode
splitting is large enough to clearly observe the key features. Panel (a) shows
the spectrum for domain 1, a strong asymmetry of the spectrum is expected
between the two K points, with only {\em two} modes carrying weight at each wave
vector. Domain 2 would have a mirror-reversed spectrum around M$_1$, again with
only \textit{two} modes visible at a general wave vector. The behavior of a
single magnetic domain of either type is clearly incompatible with the data in
Fig.~\ref{fig:110_slices_infield} (top row), which shows \textit{three} modes at
a general wave vector, with mirror symmetry of the spectrum around M$_1$.
Assuming the sample contains coexisting, equal-weight magnetic domains of both
types, the spectrum would be the sum of Figs.~\ref{fig:110_slices_chiral}(a) and
(b) plotted in Fig.~\ref{fig:110_slices_infield}(g), which restores the mirror
symmetry around M$_1$ and gives three modes at a general wave vector, as in the
data.

To test the two magnetic domains scenario further, we plot in
Fig.~\ref{fig:110_slices_infield} (bottom row) the LSWT-predicted evolution of
the spectrum as a function of field. The plotted fields were selected for best
agreement with the data in the panels above, for energy scans at the K$_1$
point. Comparison with the data shows that the key features, such as the number
of visible modes, their trend as a function of field and the overall symmetry of
the intensity pattern, are well reproduced, providing clear evidence that the
sample contains equal-weight magnetic domains of both types, as expected in the
absence of DM interactions. We attribute the remaining quantitative
discrepancies between the precise experimental dispersion shapes and the model
calculations, and the fact that the best agreement is obtained for fields
slightly different (by about 10\%) from the actual values, to quantum effects
beyond the LSWT approximation, which we have already established in
Sec.~\ref{sec:renorm} need to be included to quantitatively reproduce the
dispersions.

\section{Conclusions}
\label{sec:conclusion}

To summarize, we have reported extensive single-crystal high-resolution
inelastic neutron scattering measurements of the spin dynamics in the
pseudospin-$\half{1}$ triangular antiferromagnet \BCSO{} in the $120\ang$
ordered phase. We have observed sharp, resolution-limited magnons throughout
reciprocal space with no decay, but with a strongly renormalized dispersion and
much reduced intensities at high energies compared to linear spin wave theory.
At higher energies, we have observed a very strong continuum of magnetic
scattering extending up at least $4\times$ the maximum one-magnon energy. The
relatively large intensity in the continuum is much underestimated by linear
spin wave theory, and only some limited low-energy features are captured
qualitatively by a two-magnon cross-section, leaving unexplained a rich
structure of intensity modulations in the continuum as a function of both energy
and wave vector. We have proposed empirical wave vector-dependent
renormalizations that parametrize quantitatively the experimental dispersion in
the full three-dimensional Brillouin zone, and we have explicitly verified that
magnon decays are kinematically disallowed for the observed
strongly-renormalized dispersion, explaining why magnons are sharp throughout
the Brillouin zone. Based on a quantitative comparison of the measured
intensities with the spin wave dynamical structure factor, we have proposed that
a transfer of spectral weight occurs from the high-energy magnons (whose energy
is strongly renormalized downwards) to the higher-energy continuum. The
experimental observation of strong dispersion renormalizations and an
enhanced-intensity scattering continuum with structured intensity modulations
suggests that quantum fluctuations and interaction effects are well beyond what
can be captured by the spin wave approximation. Finally, through measurements of
the dispersion relations in $c$-axis applied magnetic field, we have determined
the presence of equal-weight magnetic domains with opposite senses for the spin
rotation in the ground state, as expected in the absence of
Dzyaloshinskii-Moriya interactions, when the sense of spin rotation in the
120$^\circ$ ordered ground state is selected via spontaneous symmetry breaking.

\begin{acknowledgments}
We thank R.~Moessner, R.~D.~Johnson, L. Balents, F.~Pollmann and R.~Verresen for
useful discussions and their interest in the work. We especially thank
C.~D.~Batista for a careful reading of the manuscript and for useful comments.
This research was partially supported by the European Research Council (ERC)
under the European Union’s Horizon 2020 research and innovation programme Grant
Agreement Number 788814 (EQFT) and by the EPSRC (UK) under Grant No.
EP/M020517/1. D.~M. acknowledges support from a doctoral studentship funded by
the EPSRC and ERC. RC acknowledges support from the National Science Foundation
under Grant No. NSF PHY-1748958 and hospitality from KITP where part of this
work was completed. The neutron scattering measurements at the ISIS Facility
were supported by a beamtime allocation from the Science and Technology
Facilities Council. In accordance with the EPSRC policy framework on research
data, access to the data will be made available from
Ref.~\onlinecite{data_archive}.

\end{acknowledgments}

\appendix

\section{Dispersion relations and dynamical structure factor in linear spin wave
theory}
\label{sec:appendix_LSWT}

This section outlines the LSWT calculation of the dispersion relation and
dynamical structure factor used in the analysis of the INS data. Based on
previous electron spin resonance,\cite{Susuki2013} nuclear magnetic
resonance,\cite{Koutroulakis2015} ultrasound velocity\cite{Quirion2015} and
neutron diffraction\cite{Ma2016} measurements of \BCSO, we assume
antiferromagnetic XXZ interactions between NN sites within the triangular layers
(intralayer exchanges) and an antiferromagnetic XXZ interaction between NN sites
on adjacent layers (interlayer exchange), as per Eq.~(\ref{eq:Hamiltonian}).
Figures~\ref{fig:Structure}(b)--(c) illustrate the exchange paths and the spin
alignments in the ground state in zero applied magnetic field. Ordered spins are
confined to the $ab$ plane, are antiparallel along $\vect{c}$, and rotate by
$120\ang$ between the three sites of every in-plane triangle. The propagation
vector for the magnetic structure is $\vect{Q}=(\third{1},\third{1},1)$. The
left/right panels in Fig.~\ref{fig:Structure}(c) show the magnetic domains with
counterclockwise/clockwise rotation.

In a magnetic field applied along the $c$ axis, the Hamiltonian in
Eq.~(\ref{eq:Hamiltonian}) acquires the additional Zeeman term
\begin{equation}
\mathcal{H}_\mathrm{Z}=-g\mu_\mathrm{B}B\sum_iS_i^z, \nonumber
\end{equation}
where $g$ is the $g$ factor and to describe the spin axes we use the Cartesian
coordinate system $(x,y,z)$ with $\vect{\hat{x}}\parallel (\vect{a}+\vect{b})$
and $\vect{\hat{z}}\parallel\vect{c}$, as illustrated in
Fig.~\ref{fig:Structure}(c) (bottom left). The magnetic structure in applied
field is a cone, where the ordered spins cant out of the $xy$ plane by an angle
$\theta$, with the in-plane components continuing to have the same pattern as in
Fig.~\ref{fig:Structure}(c). The two magnetic domains with opposite sense of
rotation in the $xy$ plane are degenerate throughout the cone phase. The canting
angle $\theta$ is obtained from minimizing the
mean-field ground state energy (per spin)
\begin{equation}
E_\mathrm{MF}=[J(\vect{Q})\cos^2{\theta}+\Delta{}J(\vect{0})\sin^2{\theta}]S^2
-g\mu_\mathrm{B}BS\sin{\theta},
\nonumber
\end{equation}
which gives
\begin{equation}
\label{eq:theta}
\sin{\theta}=\frac{g\mu_\mathrm{B}B}{2S[\Delta{}J(\vect{0})-J(\vect{Q})]}.
\nonumber
\end{equation}
Here $J(\vect{k})$ is the Fourier transform of the in-plane exchange
interactions, given by
\begin{equation}
\begin{split}
J(\vect{k})=J_1\left[\cos{2\pi{}h}+\cos{2\pi{}k}+\cos{2\pi(h+k)}\right]
+J_z\cos{\pi{}l}
\nonumber
\end{split}
\end{equation}
for a general wave vector $\vect{k}$ indexed as $(h,k,l)$ in reciprocal lattice
units of the structural unit cell, {\em i.e.}
$\vect{k}=h\vect{a}^\ast+k\vect{b}^\ast+l\vect{c}^\ast$. The canting angle
$\theta$ increases up to the saturation field $B_{\rm{sat}}=2S[\Delta
J(\vect{0})-J(\vect{Q})]/(g\mu_{\mathrm{B}})$, above which spins are entirely
polarized along the field ($\theta=\pi/2$ for $B \geq B_{\rm{sat}}$).

It is convenient to perform the analytic spin wave calculations in the cone
phase in a right-handed reference frame $(\xi,\eta,\zeta)$ that follows the
ordered spin precession in the ground state, such that $\zeta$ is along the
local ordered spin direction and $\xi$ is perpendicular to the ordered spin in
the helical plane. For concreteness, we first discuss the calculation for domain
1 with counterclockwise rotation in Fig.~\ref{fig:Structure}(c) (left panel). In
this case, the transformation from the rotating reference frame to the global
($x,y,z$) frame is obtained by first performing a rotation in the $\zeta\eta$
plane by the canting angle $\theta$, and then rotating in the $xy$ plane by the
helical angle $\phi_i=\vect{Q}\cdot\vect{r}_i+\Phi$, where $\vect{r}_i$ is the
position of the $i$th spin and $\Phi$ is the phase of the spin at the origin
[$\Phi=\half{\pi}$ for both domains illustrated in Fig.~\ref{fig:Structure}(c)].
The transformation of the spin operators is then given by
\begin{align}
S^x_i&=S^\zeta_i\cos{\theta}\cos{\phi_i}-S^\xi_i\sin{\phi_i}
-S^\eta_i\sin{\theta}\cos{\phi_i},\nonumber\\
S^y_i&=S^\zeta_i\cos{\theta}\sin{\phi_i}+S^\xi_i\cos{\phi_i}
-S^\eta_i\sin{\theta}\sin{\phi_i},\nonumber\\
S^z_i&=S^\zeta_i\sin{\theta}+S^\eta_i\cos{\theta}.\nonumber
\end{align}
The spin Hamiltonian for the NN intralayer interactions [first term of
Eq.~(\ref{eq:Hamiltonian})] in the rotating reference frame has the form
\begin{align}
\mathcal{H}_\mathrm{NN}=&J_1\sum_{\langle{ij}\rangle}
[(\cos^2{\theta}\cos{\phi_{ij}}
+\Delta\sin^2{\theta})S^\zeta_iS^\zeta_j\nonumber\\
&+\cos{\phi_{ij}}S^\xi_iS^\xi_j
+(\sin^2{\theta}\cos{\phi_{ij}}
+\Delta\cos^2{\theta})S^\eta_iS^\eta_j\nonumber\\
&+\sin{\theta}\sin{\phi_{ij}}(S^\xi_iS^\eta_j-S^\eta_iS^\xi_j)]\nonumber,
\end{align}
where $\phi_{ij}=\phi_i-\phi_j$. A similar expression describes the interlayer
interactions. The advantage of working in the rotating frame is that all spins
are ferromagnetically aligned and the calculation is reduced to one magnetic
sublattice and a reduced hexagonal unit cell $a\times b \times (c/2)$.

Using a Holstein-Primakoff transformation,\cite{Holstein1940} a Fourier
transformation, and neglecting terms higher than quadratic order in the boson
operators, the spin Hamiltonian in the rotating frame is obtained
as\cite{Veillette}
\begin{align}
\mathcal{H}=&\frac{1}{2}\sum_{\vect{k}}\matr{X}^\dagger\matr{H}\matr{X}
-Ng\mu_\mathrm{B}B(S+1/2)\sin{\theta}\nonumber\\
&+N[J(\vect{Q})\cos^2{\theta}+\Delta{}J(\vect{0})\sin^2{\theta}]S(S+1)\nonumber
\end{align}
where the sum is over all wave vectors $\vect{k}$ in the first Brillouin zone of
the reduced unit cell and $N$ is the total number of spin sites. The operator
basis is chosen to be
$\matr{X}^\dagger=
\begin{pmatrix}\alpha_{\vect{k}}^\dagger&\alpha_{-\vect{k}}\end{pmatrix}$,
where $\alpha_{\vect{k}}^\dagger$ ($\alpha_{\vect{k}}$) creates (annihilates) a
plane-wave magnon. The Hamiltonian matrix then has the form
\begin{equation}
\matr{H}=
\begin{pmatrix}
A_{\vect{k}}+C_{\vect{k}}&B_{\vect{k}}\\
B_{\vect{k}}&A_{\vect{k}}-C_{\vect{k}}\\
\end{pmatrix},
\nonumber
\end{equation}
where
\begin{align}
A_{\vect{k}}&=S(a_{\vect{k}}+b_{\vect{k}}),\nonumber\\
B_{\vect{k}}&=S(a_{\vect{k}}-b_{\vect{k}}),\nonumber\\
C_{\vect{k}}&=S\sin{\theta}\left[J(\vect{k}+\vect{Q})-J(\vect{k}
-\vect{Q})\right],
\label{eq:Ck}
\end{align}
and
\begin{align}
a_{\vect{k}}&=\left[\Delta{}J(\vect{k})-J(\vect{Q})\right]\cos^2{\theta}
+b_{\vect{k}}\sin^2{\theta},\nonumber\\
b_{\vect{k}}&=\half{1}\left[J(\vect{k}-\vect{Q})+J(\vect{k}+\vect{Q})\right]
-J(\vect{Q}).\nonumber
\end{align}

Using standard methods to diagonalize the bilinear boson
Hamiltonian,\cite{White1965} the dispersion relation is obtained as
\begin{equation}
\hbar\omega(\vect{k})=\sqrt{A_{\vect{k}}^2-B_{\vect{k}}^2}+C_{\vect{k}},
\label{eq:Dispersion}
\end{equation}
which by periodicity holds for a general wave vector $\vect{k}$ in reciprocal
space. The one-magnon excitations are polarized transverse to the ordered spin
direction $\zeta$, and the dynamical structure factors (per spin) are obtained
as
\begin{align}
S^{\xi\xi}(\vect{k},\omega)&=\frac{Z_\xi{}S}{2}(u_{\vect{k}}
+v_{\vect{k}})^2\delta[\hbar\omega-\hbar\omega(\vect{k})]\nonumber\\
&=\frac{Z_\xi{}S}{2}\frac{A_{\vect{k}}
+B_{\vect{k}}}{\hbar\omega(\vect{k})}\delta[\hbar\omega
-\hbar\omega(\vect{k})],\label{eq:S_xixi}\\
S^{\eta\eta}(\vect{k},\omega)&=\frac{Z_\eta{}S}{2}(u_{\vect{k}}
-v_{\vect{k}})^2\delta[\hbar\omega-\hbar\omega(\vect{k})]\nonumber\\
&=\frac{Z_\eta{}S}{2}\frac{A_{\vect{k}}-B_{\vect{k}}}{\hbar\omega(\vect{k})}
\delta[\hbar\omega-\hbar\omega(\vect{k})],\label{eq:S_etaeta}\\
S^{\xi\eta}(\vect{k},\omega)&=-S^{\eta\xi}(\vect{k},\omega)=
i\frac{S}{2}\delta[\hbar\omega-\hbar\omega(\vect{k})],\label{eq:S_xieta}
\end{align}
where $u_{\vect{k}}=\cosh{\Theta_{\vect{k}}}$,
$v_{\vect{k}}=\sinh{\Theta_{\vect{k}}}$ and
$\tanh{2\Theta_{\vect{k}}}=B_{\vect{k}}/A_{\vect{k}}$. The intensity prefactors
for in-plane ($Z_\xi$) and out-of-plane magnons ($Z_\eta$) are both unity in
LSWT; they are introduced here as a way to parametrize an intensity
renormalization due to effects beyond the LSWT approximation.

The two-magnon (${2\mathcal{M}}$) excitations are polarized longitudinal to the
spin direction, and the dynamical structure factor (per spin) is obtained as
\begin{equation}
\begin{split}
S^{\zeta\zeta}_{2\mathcal{M}}(\vect{k},\omega)=
\frac{Z_\zeta}{2N}\sum_{\vect{k}_1,\vect{k}_2}(u_{-\vect{k}_1}v_{\vect{k}_2}
+u_{\vect{k}_2}v_{-\vect{k}_1})^2\\
\times\delta[\hbar\omega-\hbar\omega(\vect{k}_1)-\hbar\omega(\vect{k}_2)]
\delta(\vect{k}+\vect{k}_1-\vect{k}_2+\vect{\tau}),
\label{eq:S2m}
\end{split}
\end{equation}
with the density of states (meV$^{-1}$Co$^{-1}$) for two-magnon excitations
given by
\begin{equation}
D(\vect{k},\omega)=\frac{1}{N}\sum_{\vect{k}_1,\vect{k}_2}\delta[\hbar\omega
-\hbar\omega(\vect{k}_1)-\hbar\omega(\vect{k}_2)]\delta(\vect{k}+\vect{k}_1
-\vect{k}_2+\vect{\tau}),\nonumber
\end{equation}
where $\vect{\tau}$ is a reciprocal lattice vector of the reduced unit cell.
Similar to the expressions for the one-magnon dynamical structure factor, we
introduce in Eq.~(\ref{eq:S2m}) a two-magnon intensity prefactor $Z_\zeta$ to
parametrize an intensity renormalization attributed to effects beyond the LSWT
approximation. Following Ref.~\onlinecite{Mourigal2013}, we neglect the mixed
transverse-longitudinal correlations (such as $S^{\xi\zeta}$), as for the TLHAF
model they have relatively negligible weight compared to the purely transverse
or purely longitudinal correlations.

For a spin-$\half{1}$ system, the dynamical structure factor components are
required to satisfy the sum rule\cite{Lorenzana2005}
\begin{equation}
\frac{1}{N}\sum_{\vect{k}}\int^{\infty}_{-\infty}d(\hbar\omega)\;
S^{\alpha\alpha}(\vect{k},\omega)=\frac{1}{4},\nonumber
\label{eq:sum_rules}
\end{equation}
where $\alpha=\xi,\eta,\zeta$ and the sum is over all wave vectors in the first
Brillouin zone of the reduced unit cell. Note that the longitudinal component
$S^{\zeta\zeta}$ includes two-magnon scattering $S^{\zeta\zeta}_{2\mathcal{M}}$
and elastic Bragg scattering $(S-\Delta S)^2$, which gives the following sum
rule for the two-magnon contribution:
\begin{equation}
\frac{1}{N}\sum_{\vect{k}}\int^{\infty}_{-\infty}d(\hbar\omega)\;
S^{\zeta\zeta}_{2\mathcal{M}}(\vect{k},\omega)=
\frac{1}{4}-(S-\Delta{S})^2,\nonumber
\end{equation}
where $\Delta{S}$ is the reduction in the ordered spin moment in the ground
state due to zero-point spin wave fluctuations. For the Hamiltonian parameters
in Table~\ref{tab:pars}, $\Delta{S}=0.152$ and the above sum rules are satisfied
for intensity prefactor values $Z_\xi=0.804$, $Z_{\eta}=0.734$ and
$Z_{\zeta}=0.733$. In the fits to the experimental spin wave data, we allowed
the relative intensity of in-plane to out-of-plane magnons to vary
unconstrained, with the best overall agreement found for $Z_\eta/Z_{\xi}=0.60$,
to be compared with 0.91 imposed by the sum rule constraints and 1 in LSWT. In
all calculations, the two-magnon scattering intensity was scaled to the in-plane
one-magnon intensity assuming both satisfy the sum rule constraints, which gives
$Z_\zeta/Z_{\xi}=0.91$.

Rotating back to the fixed global frame, the dynamical structure factors for
one-magnon ($1\mathcal{M}$) and two-magnon ($2\mathcal{M}$) excitations are as
follows:
\begin{align}
S^{zz}_{1\mathcal{M}}(\vect{k},\omega)=&
S^{\eta\eta}(\vect{k},\omega)\cos^2{\theta},\nonumber\\
S^{zz}_{2\mathcal{M}}(\vect{k},\omega)=&
S^{\zeta\zeta}_{2\mathcal{M}}(\vect{k},\omega)\sin^2{\theta},\nonumber\\
S^{xx}_{1\mathcal{M}}(\vect{k},\omega)=&
\frac{1}{4}[S^{\xi\xi}(\vect{k}-\vect{Q},\omega)
+S^{\xi\xi}(\vect{k}+\vect{Q},\omega)]\nonumber\\
&+ \frac{1}{4}\sin^2{\theta}[S^{\eta\eta}(\vect{k}-\vect{Q},\omega)
+S^{\eta\eta}(\vect{k}+\vect{Q},\omega)]\nonumber\\
&+ \frac{i}{2}\sin{\theta}[S^{\xi\eta}(\vect{k}+\vect{Q},\omega)
-S^{\xi\eta}(\vect{k}-\vect{Q},\omega)],\nonumber\\
S^{xx}_{2\mathcal{M}}(\vect{k},\omega)=&
\frac{1}{4}\cos^2{\theta}[S^{\zeta\zeta}_{2\mathcal{M}}(\vect{k}
-\vect{Q},\omega)+S^{\zeta\zeta}_{2\mathcal{M}}(\vect{k}+\vect{Q},\omega)],
\nonumber
\end{align}
where by symmetry $S^{xx}(\vect{k},\omega)=S^{yy}(\vect{k},\omega)$. The
two-magnon density of states in the global frame is obtained as
\begin{equation}
\widetilde{D}(\vect{k},\omega)=
\frac{1}{2}\left[D(\vect{k}-\vect{Q},\omega)+D(\vect{k}+\vect{Q},\omega)\right].
\label{eq:DoS}
\end{equation}
The above analytic expressions for the dispersion relations and dynamical
structure factor were checked explicitly against  numerical calculations
performed using \textsc{spinw}.\cite{Toth2015} The interpretation of the above
equations for the one-magnon dynamical structure factor is that in the global
frame there are two in-plane-polarized modes
$\omega(\vect{k}-\vect{Q})\equiv\omega^{-}(\vect{k})$ and
$\omega(\vect{k}+\vect{Q})\equiv\omega^{+}(\vect{k})$ and one out-of-plane mode
$\omega(\vect{k})$, so there are three dispersion branches for a general wave
vector $\vect{k}$.

All the above expressions for the dispersion relation and dynamical structure
factor are for the magnetic domain 1 in Fig.~\ref{fig:Structure}(c) (left
panel); the results for domain 2 (right panel) are obtained by replacing
$\vect{Q}$ with $-\vect{Q}$ in Eq.~(\ref{eq:Ck}), which changes the sign of the
$C_{\vect{k}}$ term in Eq.~(\ref{eq:Dispersion}) with the $A_{\vect{k}}$ and
$B_{\vect{k}}$ terms unchanged. This implies that in zero field when $\theta=0$
and $C_{\vect{k}}=0$, the two domains have identical dispersions and dynamical
structure factors, and so cannot be distinguished experimentally. However, for
finite field $B$ the $C_{\vect{k}}$ term is finite and the two domains have
different primary mode dispersions. This is illustrated in
Fig.~\ref{fig:110_slices_chiral}, which shows the calculated spin wave spectrum
in finite field [panels (a) and (b) for domains 1 and 2, respectively], showing
that the primary magnon dispersion (magenta solid line) is different in the two
cases, with soft modes at different wave vectors [K$_1$ in (a) and K$_3$ in
(b)].

Finally, total neutron scattering cross-section including the neutron
polarization factor is
\begin{align}
I(\vect{k},\omega)=&Z[n(\hbar\omega)+1]f^2(\abs{\vect{k}})\nonumber\\
&\times\sum_{\alpha}\left(1-\frac{k_\alpha^2}{\abs{\vect{k}}^2}\right)
S^{\alpha\alpha}(\vect{k},\omega),
\label{eq:Intensity}
\end{align}
where $Z$ is an overall intensity scale factor, 
$n(\hbar\omega)=1/\left(e^{\hbar\omega/k_\mathrm{B}T}-1\right)$ is the
finite-temperature Bose factor, $f(\abs{\vect{k}})$ is the spherical magnetic
form factor for Co$^{2+}$ ions, and $k_\alpha$ denotes the $\alpha=x,y,z$
component of the wave vector transfer $\vect{k}$.

In the fits to the experimental data, we used the renormalized dispersion (see
Appendix~\ref{sec:appendix_renorm}) in place of $\hbar\omega(\vect{k})$ in the
dynamical structure factor expressions in
Eqs.~(\ref{eq:S_xixi})--(\ref{eq:S_xieta}). The effects of the instrumental
energy resolution were included by replacing the delta functions in the same
equations with a lineshape of finite energy width that could describe well the
observed profile of the incoherent elastic line. For each separate instrument
configuration, the appropriate energy resolution lineshape was parametrized by a
main Gaussian with an additional less intense Gaussian on the low-energy side to
reproduce the observed slightly-asymmetric energy lineshape. In the fits, the
resolution profile was assumed constant as a function of energy transfer.

\section{Empirical renormalizations of the LSWT dispersion}
\label{sec:appendix_renorm}

\begin{table}
\centering
\begin{tabular}{c d l}
\hline\hline
Parameter & \multicolumn{2}{c}{Value} \\
\hline	
$J_1$ & 1.653 & meV \\
$J_z$ & 0.082 & meV \\
$\Delta$ & 0.949 & \\
$Z_\eta/Z_\xi$ & 0.60 & \\ 
$\alpha_\mathrm{M}$ & 1.434 & meV \\
$\beta_\mathrm{M}$ & 0.019 & meV \\
$\gamma_\mathrm{M}$ & 17.18 & meV\AA$^2$ \\ 
$\delta_\mathrm{M}$ & 0.151 & \\
$\kappa_\mathrm{M}$ & 1.333  & \\
$\alpha_\mathrm{K/2}$ & 2.040 & meV\\
$\beta_\mathrm{K/2}$ & -0.025  & meV \\
$\gamma_\mathrm{K/2}$ &  13.12 & meV\AA$^2$ \\ 
$\delta_\mathrm{K/2}$ &  0.192 & \\	
$\kappa_\mathrm{K/2}$ & 1^{\dag} & \\
$\epsilon$ & 0.1566^{\dag} & \\
$Z_\zeta/Z_\xi$ & 0.91^{\dag} & \\ 
\hline \hline
\end{tabular}
\caption{Parameter values for the best fit to the observed one-magnon dispersion
relations, obtained from a global fit to several selected scans through the
four-dimensional INS data. In the fit the Hamiltonian parameters
($J_1$,$J_z$,$\Delta$) were constrained to reproduce the saturation
magnetization field as per Eq.~(\ref{eq:constraint}). The dagger $^{\dag}$
indicates parameter values kept fixed in the global fit. The table omits the
overall intensity scale $Z$, the parametrization of the non-magnetic background
and the instrumental energy resolution, as these vary between different
measurement configurations and are discussed elsewhere. For descriptions of the
listed parameters, see Appendices~\ref{sec:appendix_LSWT} and
\ref{sec:appendix_renorm}. \label{tab:pars}}
\end{table}

In this section, we detail the empirical renormalizations applied to the
analytic LSWT dispersion relation in order to fit the experimental magnon
dispersion. In particular we consider the introduction of soft modes in the
dispersion at the M and near K/2 points.

To introduce local minima in the dispersion, we consider the virtual mixing of
the bare dispersion $\hbar\omega_\mathrm{LSWT}$ in Eq.~(\ref{eq:Dispersion})
with fictitious gapped parabolic modes $\hbar\Omega_i$, centered near wave
vector positions $i=$ M and K/2. This mixing can be parametrized in the basis of
the two modes by a $2\times2$ Hamiltonian matrix
\begin{equation}
\matr{H}=
\begin{pmatrix}
\hbar\omega_\mathrm{LSWT}&c_i\\
c_i&\hbar\Omega_i\\
\end{pmatrix},
\nonumber
\end{equation}
where the off-diagonal coupling term is defined as
$c_i\equiv\delta_i\,\hbar\omega_\mathrm{LSWT}$, with $\delta_i$ a dimensionless
parameter. This form ensures the coupling $c_i$ is largest near the top of the
dispersion and becomes negligibly small at low energies. The above Hamiltonian
can then be diagonalized to obtain the eigenenergies
\begin{equation}
\lambda^\pm=\frac{\hbar\omega_\mathrm{LSWT}+\hbar\Omega_i}{2}\pm
\sqrt{\left(\frac{\hbar\omega_\mathrm{LSWT}-\hbar\Omega_i}{2}\right)^2+c_i^2},
\nonumber
\end{equation}
where the lower mode $\lambda^-$ is a smoothly-varying function that inherits a
local minimum from the gapped virtual mode $\hbar\Omega_i$ and interpolates
towards the unperturbed $\hbar\omega_\mathrm{LSWT}$ in the regions away from the
soft mode. This is graphically illustrated in Fig.~\ref{fig:renorm}, compare the
solid magenta line ($\lambda^-$) with the dashed magenta line
($\hbar\omega_\mathrm{LSWT}$).

The fictitious gapped modes were parametrized by the general dispersion form
\begin{equation}
\hbar\Omega_i=\alpha_i+\beta_i\cos{\pi{}l}+\gamma_i[(q_x-q_{ix})^2
+\kappa_i(q_y-q_{iy})^2],
\nonumber
\end{equation}
where the first term ($\alpha_i$) parametrizes the overall energy gap, the
second term allows for a dispersion along the interlayer $l$ direction, and
$\gamma_i$ is the coefficient of the in-plane quadratic dispersion.
$(q_{ix},q_{iy})$ are the in-plane wave vector coordinates (in \AA$^{-1}$) of
the paraboloid center (minimum energy gap) in a Cartesian reference frame, where
the $q_y$ coordinate is along the direction from the closest $\Gamma$ point to
the paraboloid center and $q_x$ is transverse to $q_y$ in the $hk$ plane.
$\kappa_i$ parametrizes the relative dispersions along the two orthogonal
in-plane directions, \textit{i.e.} $\kappa_i=1$ corresponds to an isotropic
dispersion with circular constant-energy contours and $\kappa_i>1$ corresponds
to elliptical constant-energy contours elongated along the transverse $q_x$
direction. Figure~\ref{fig:hkslices}(g) shows clear oval-shaped contours around
the M points, elongated along the hexagonal zone-boundary contour (dashed white
line), and this elongation was parametrized in the fit by the ellipticity
parameter $\kappa_\mathrm{M}\simeq1.3$. For the soft modes near K/2, we found an
isotropic description to be sufficient, so we fixed $\kappa_\mathrm{K/2}=1$ in
the fit. As expected for a quasi-2D system, the fitted interlayer dispersion is
almost negligible at the relatively high energies of the soft modes
($\abs{\beta_i}/\alpha_i\simeq0.01$ for both $i=\mathrm{M}$ and K/2).

The procedure for obtaining the renormalized magnon dispersion $\hbar\omega$
from the bare $\hbar\omega_\mathrm{LSWT}$ after considering the couplings with
both types of virtual parabolic modes is illustrated in Fig.~\ref{fig:renorm},
where the lower (magenta) solid line has the desired soft modes at both types of
positions with symmetric dispersions around the local minima, as seen in the
experimental data. Note that the original magnon dispersion is not perfectly
sinusoidal along the $\Gamma$-K line, with the maximum slightly offset from the
halfway position K/2; so in order to obtain an approximately symmetric shape for
the dispersion near the soft mode along the $\Gamma$-K direction, the paraboloid
$\hbar\Omega_\mathrm{K}/2$ was centered at position
$\epsilon(\vect{a}^\ast+\vect{b}^\ast)$, with $\epsilon$ slightly offset (see
Table~\ref{tab:pars}) from the value 1/6 that corresponds to the exact K/2 wave
vector position. Using the latter position would have resulted in a highly
asymmetric shape of the dispersion near the upper soft mode, not compatible with
the experimental data in Fig.~\ref{fig:KMK_3meV_7meV}.

In order to calculate the renormalized dispersion relation, it is sufficient to
work in the minimal Brillouin zone sector $\Gamma$-M-K-$\Gamma$ in the $hk$
plane and $0\leq l<1$, as any general wave vector $\vect{k}$ can be remapped to
this volume using reciprocal lattice translations followed by symmetry
operations of the $6/mmm$ lattice point group. For wave vectors within this
minimal reciprocal space volume, we calculated iteratively the mixing of
$\hbar\omega_\mathrm{LSWT}$ with virtual paraboloids located at equivalent (up
to reciprocal lattice translations or lattice point group symmetry operations) M
and K/2-type positions within a large radius in the two-dimensional reciprocal
space at the same $l$ value; in this way, we ensured the ``final'' renormalized
dispersion (that is fitted to the experimental data) still satisfies all lattice
point group symmetries and is numerically smooth (so there is no step change in
gradient across the minimal volume boundaries). A contour map of the
renormalized dispersion surface in the $(hk0)$ plane is shown in
Fig.~\ref{fig:map_dispersion}(b).

The Hamiltonian and dispersion renormalization parameters obtained from a best
fit to the experimental data are listed in Table~\ref{tab:pars}, and numerical
code to generate the dispersion relation from these parameters is available from
Ref.~\onlinecite{data_archive}.

We note that in order to capture all modulations of the full magnon dispersion
surface in a transparent way that can also be easily implemented analytically,
several empirical parameters have been introduced: three Hamiltonian parameters
($J_1$,$J_z$,$\Delta$), five parameters ($\alpha_i$, $\beta_i$, $\gamma_i$,
$\delta_i$, $\kappa_i$) for each of the soft modes at M and near K/2, in
addition to independent intensity scale factors for the in-plane and
out-of-plane magnons. Although some parameters were kept fixed in the fit and
additional constraints were imposed, this still left a very large number of
degrees of freedom in the fit (over 10) and in practice many parameters were
strongly correlated. Therefore, Table~\ref{tab:pars} parameter values are to be
interpreted as {\em representative} values for the best level of agreement that
can be obtained with the data; the meaningful result of the analysis is the
final parametrized dispersion surface obtained with those parameters and its
specific features, not the individual values of each of the parameters.

\end{document}